\documentclass[apj]{emulateapj}

\usepackage{graphicx}	
\usepackage{amsmath}	
\usepackage{amssymb}	
\usepackage{xcolor}      
\usepackage{hyperref}

\newcommand{\fpath}{.}
\renewcommand{\vec}[1]{\boldsymbol{#1}}
\newcommand{\hvec}[1]{\hat{\boldsymbol{#1}}}
\newcommand{\pd}[2]{\frac{\partial #1}{\partial #2} }
\newcommand{\HALF}{\frac{1}{2}}

\newcommand{\jump}[1]{\left[ #1 \right]}

\newcommand{\tmop}[1]{\ensuremath{\operatorname{#1}}}

\newcommand{\blue}{\color{black}}
\newcommand{\black}{\color{black}}

\newcommand{\av}[1]{\left< {#1} \right>}
\newcommand{\DS}{\displaystyle}
\newcommand{\tens}[1]{\mathsf{#1}}
\newcommand{\p}{{\mathrm{p}}}
\newcommand{\e}{\mathrm{e}}
\newcommand{\vsh}{v_{\rm sh}}
\newcommand{\CMB}{{\textsc{\tiny CMB}}}

\begin{document}

\shorttitle{Modeling Non-thermal emission from Relativistic Magnetized flows.}
\title{{\blue A Particle Module for the PLUTO code: II - Hybrid Framework for Modeling Non-thermal emission from Relativistic Magnetized flows.}}

\shortauthors{Vaidya et. al}

\author{Bhargav Vaidya}
\affiliation{Centre of Astronomy, Indian Institute of Technology Indore, Khandwa Road, Simrol , Indore 453552, India}
\affiliation{Dipartimento di Fisica, University of Torino, via Pietro Giuria 1, I-10125 Torino, Italy}

\author{Andrea Mignone}
\affiliation{Dipartimento di Fisica, University of Torino, via Pietro Giuria 1, I-10125 Torino, Italy}

\author{Gianluigi Bodo}
\affiliation{INAF, Osservatorio Astrofisico di Torino, Strada Osservatorio 20, I-10025 Pino Torinese, Italy }

\author{Paola Rossi}
\affiliation{INAF, Osservatorio Astrofisico di Torino, Strada Osservatorio 20, I-10025 Pino Torinese, Italy }

\author{Silvano Massaglia}
\affiliation{Dipartimento di Fisica, University of Torino, via Pietro Giuria 1, I-10125 Torino, Italy}

\email{bvaidya@iiti.ac.in}

\begin{abstract}
We describe a new hybrid framework to model non-thermal spectral signatures from highly energetic particles embedded in a large-scale classical or relativistic MHD flow.
Our method makes use of \textit{Lagrangian} particles moving through an Eulerian grid where the (relativistic) MHD equations are solved concurrently.
Lagrangian particles follow fluid streamlines and represent ensembles of (real) relativistic particles with a finite energy distribution.
The spectral distribution of each particle is updated in time by solving the relativistic cosmic ray transport equation based on local fluid conditions.
This enables us to account for a number of physical processes, such as adiabatic expansion, synchrotron and inverse Compton emission.
An accurate semi-analytically numerical scheme that combines the method of characteristics with a Lagrangian discretization in the energy coordinate is described.

In presence of (relativistic) magnetized shocks, a novel approach to consistently model particle energization due to diffusive shock acceleration has been presented.
Our approach relies on a refined shock-detection algorithm and updates the particle energy distribution based on the shock compression ratio, magnetic field orientation and amount of (parameterized) turbulence.
The evolved  distribution from each \textit{Lagrangian} particle is further used to produce observational signatures like emission maps and polarization signals accounting for proper relativistic corrections.
We further demonstrate the validity of this hybrid framework using standard numerical benchmarks and evaluate the applicability of such a tool to study high energy emission from extra-galactic jets. 
\end{abstract}




\section{Introduction}
%
%
%

Magnetized and relativistic large scale flows in the form of jets are a common observational feature seen for example in active galactic nuclei (AGNs), Gamma-ray bursts and micro-quasars. 
The dominant emission is originated by non-thermal processes from high energy particles. 
Multi-wavelength observations covering a wide spectrum from Radio wavelengths to TeV Gamma ray emission provides valuable insights into the micro-physical processes that occur in jets and lead to the observed radiation. 
The length scales associated with these micro-physical processes are many orders of magnitude smaller than the physical jet scales that can range up to few tens of kilo-parsec.
Connecting a bridge between these scales poses a serious challenge to theoretical modeling of the emission from AGN jets. 
In the present work, we aim to build a quantitative connection between such disjoint scales by developing a numerical tool that could simulate multi-dimensional flow pattern treating small-scale processes in a sub-grid manner. 
In this work, we describe such a tool that consistently accounts for most of the micro-physical processes.

The general analytical picture of multi-wavelength radiation from beamed relativistic magnetized jet was proposed by works in the eighties \citep[e.g.][]{Blandford:1979, Marscher:1980, Konigl:1981}. 
Since then, synchrotron emission signatures from large scale jets are obtained from time-dependent simulations through post-processing. 
In the relativistic hydrodynamic context, transfer functions between thermal and non-thermal electrons in jet are used \citep{Gomez:1995, Gomez:1997, Aloy:2000} whereas in case of relativistic MHD calculations, the magnetic structure inside the jet is used to compute synchrotron emission maps \citep[e.g.][]{Porth:2011, Hardcastle:2014, English:2016}.

  {\blue Formalism to study the micro-physics of particle acceleration at shocks using hybrid implementations combining both particles and grid-based fluid descriptions have also been developed targeting different scales of interest.
At the scales of electron's gyro radius, the most consistent approach is that of Particle in Cell (PIC). 
Several groups have applied this kinetic approach to understand shock acceleration at relativistic shocks \citep[e.g.,][and references therein]{Sironi:2015}. 
A hybrid MHD-PIC approach can be used to study the shock acceleration phenomenon on slightly large length scales typically of the order of few thousands of proton gyro-scales. 
Such an approach developed by \citep[e.g.,][]{Bai:2015, vanMarle:2017, Mignone:2018} describes the interaction between collisionless cosmic ray particles and a thermal plasma. 
Similarly, \cite{Daldorff:2014} proposed a hybrid approach for the BATS-R-US code that combines Hall-MHD and PIC methods in order to capture small-scale kinetic effects in magnetosphere simulations.}

An alternative approach in  numerical modeling of non-thermal emission from astrophysical jet treats the population of non-thermal electrons as separate particle entities suspended in fluid.
Effects due to synchrotron aging in presence of shock acceleration under the test particle limit were studied for radio galaxies using multi-dimensional classical MHD simulations by \citep{Jones:1999, Tregillis:2001}.  
Acceleration of test particles and subsequent radiative losses in presence of shocks formed via hydro-dynamic Kelvin Helmholtz vortices were studied by \cite{Micono:1999}. 
Such an hybrid framework of combining test particles with classical fluid has also been used effectively to study cosmic-ray transport in cosmological context \citep{Miniati:2001}. 
For relativistic hydrodynamic flows, populations of non-thermal particles (NTPs) have been included to study non-thermal emission from internal shocks in Blazars \citep{Mimica:2009, Mimica:2012, Fromm:2016}. 
Recent relativistic hydro-dynamical simulations using NTPs have also been applied for a study of star-jet interactions in AGNs \citep{DeLaCita:2016}. 
There are two most critical limitations with above models using NTPs. 
Firstly as the fluid simulations are done with RHD, magnetic field strengths are assumed to be in equipartition with the internal energy density. 
This ad-hoc parameterized assumption of magnetic field strengths can affect the estimation of the spectral break in the particle distribution due to synchrotron processes. 
The second simplifying assumption in these models is the choice of a constant value for the power law index $\mathcal{N}(E) \propto E^{-m}$, ($m = 2.0$ \citep{DeLaCita:2016} and $m = 2.23$ \citep{Fromm:2016}) in the recipe of particle injection at shocks.

In the present work, we describe methods used to overcome the above limitations with an aim to build a state-of-the-art hybrid framework of particle transport  to model high energy non-thermal emission from large scale 3D RMHD simulations. 
Our sub-grid model for shock acceleration incorporates the dependence of the spectral index on the shock strength and magnetic field orientation. 
The magnetic fields obtained from our RMHD simulations are used to compute radiation losses due to synchrotron and Inverse Compton (IC) emission in a more accurate manner without any assumption on equipartition.
Further, we also incorporate the effects of relativistic aberration in estimating of the polarized emission due to synchrotron processes. 

\blue
Unlike the MHD-PIC approach \citep[e.g.][]{Bai:2015, vanMarle:2017, Mignone:2018}, 
we do not consider the feed-back of motion of particles on the fluid.
This surely does not allow us to study the associated non-linear coupling effects. 
However, we can certainly extend the applicability of the presented hybrid framework, to observable scales, whereby micro-physical aspects of spectral evolution are treated using sub-grid physics based on local fluid conditions.
Further, we have only considered the spectral evolution for electrons in presence of magnetic fields and shocks. Modifications in the mass and time scales would be required if the physics of acceleration of protons are to be incorporated.  
Also, the protons are expected to suffer from an negligible amount of synchrotron loss in comparison with electrons which would significantly reflect in the spectral behavior of high energy protons in comparison to electrons. 
Additionally, the post-shock spectral evolution is different for protons as demonstrated by PIC simulations for all kinds of shocks \cite[e.g.][]{Sironi:2013aa, Park:2015, Marcowidth:2016}.
This sub-grid physics associated with acceleration and radiative loss of protons is not included in the present work. 
\black

The paper is organized as follows - detailed theoretical description used for our hybrid particle \& fluid framework is described in Sec.~\ref{sec:numframe}. In particular, the transport equation for particle spectral evolution is given in Sec.~\ref{sec:cr_transeq}, details of numerical implementation are outlined in Sec.~\ref{sec:lp}, different micro-physical processes considered are elaborated in details in Sec.~\ref{ssec:radloss} and ~\ref{ssec:dsa}. The post-processing methods used to obtain emissivity and polarization signatures from particles are described in Sec~\ref{ssec:postproc}.
In Sec.~\ref{sec:numtests}, we demonstrate the accuracy of the developed hybrid framework using standard tests and further go on to describe the astrophysical applications in Sec.~\ref{sec:application}.

\section{Numerical Framework}
\label{sec:numframe}
%
%
%

\subsection{The Cosmic Ray Transport Equation}
\label{sec:cr_transeq}
%

The transport equation for cosmic rays in a scattering medium has been derived, in the classical case, by several authors \citep[see e.g.][]{Parker:1965, Jokipii:1970, Skilling:1975, Webb:1979} and, in the relativistic case, by \cite{Webb:1989}.
Let $f_0(x^\mu, p)$ be the isotropic distribution function of the non-thermal particle in phase space, where $x^\mu$ and $p$ denote the position four-vector and the momentum magnitude, respectively; the transport equation then reads \citep{Webb:1989}
\begin{equation}\label{eq:CR_full}
\begin{split}
    \nabla_{\mu}\left(u^{\mu}f_0 + q^{\mu}\right)
  + \frac{1}{p^2} \frac{\partial}{\partial p}\Big[
  - \frac{p^3}{3} f_0 \nabla_{\mu}u^{\mu} + \av{\dot{p}}_{l} f_0  & \\
  - \Gamma_{\rm visc}p^{4} \tau \frac{\partial f_0}{\partial p} - p^{2}D_{pp}
  \frac{\partial f_0}{\partial p} - p (p^0)^{2} \dot{u}_{\mu} q^{\mu}
 \Big] = 0
\end{split}
\end{equation}
where the terms in round brackets describe particle transport by convection, and particle transport by diffusion, respectively.
Here $u^\mu$ is the bulk four-velocity of the surrounding fluid while $q^\mu$ is the spatial diffusion flux.
The terms in square bracket are responsible for evolution in momentum space and describe, respectively:
\begin{itemize}
  \item the energy changes due to adiabatic expansion;
  \item the losses associated with synchrotron and IC emission
        (here $\av{\dot{p}}_l$ is the average momentum change due to
         non-thermal radiation), see Sec. \ref{ssec:radloss};
  \item the acceleration term due to fluid shear, where $\Gamma_{\rm visc}$
        is the shear viscosity coefficient;
  \item Fermi II order process, where $D_{pp}$ is the diffusion coefficient
        in momentum space;
  \item non-inertial energy changes associated with the fact that particle
        momentum $p$ is measured relative to a local Lorentz frame moving
        with the fluid (here $p^0$ is the temporal component of the momentum
        four-vector while $\dot{u}_\mu$ is the four-acceleration).
\end{itemize}

For the present purpose, we shall neglect particle transport due to spatial diffusion, (i.e., $q^{\mu} = 0$) and, for simplicity, ignore particle energization due to shear ($\Gamma_{\rm visc} = 0$), Fermi second order processes ($D_{pp} = 0$) and the last term involving non-inertial energy changes (as $q^{\mu} = 0$). 
Eq. (\ref{eq:CR_full}) then reduces to 
\begin{equation}
  \nabla_{\mu}(u^{\mu}f_0)
+ \frac{1}{p^2} \frac{\partial}{\partial p}\left[
- \frac{p^3}{3} f_0 \nabla_{\mu}u^{\mu}  + \av{\dot{p}}_{l} f_0
  \right] = 0.
\end{equation}
On expanding the derivative in the first term and using the fact that,
\begin{equation}
  u^{\mu}\nabla_{\mu} =
  \gamma\left(\frac{\partial}{\partial t} + v^{i}\frac{\partial}{\partial x_{i}}\right)
  \equiv \frac{d}{d\tau}
\end{equation}
is the Lagrangian derivative with respect to proper time, related  to the laboratory time by  $d\tau = dt/\gamma$, where $\gamma$ is the bulk Lorentz factor,  we obtain
\begin{equation}\label{eq:CR_p}
p^{2} \frac{df_0}{d\tau} + \frac{\partial}{\partial p}
 \left[ - \frac{p^3}{3} f_0 \nabla_{\mu}u^{\mu}
        + \av{\dot{p}}_{l} f_0\right] = -p^{2}f_0\nabla_{\mu}u^{\mu}
\end{equation}
We now define ${\cal N}(p, \tau) = \int d\Omega p^{2} f_0 \approx 4\pi p^{2}f_0$,  taking into account the assumption of  isotropy for distribution of particles in momentum space.
Physically, ${\cal N}(p, t)dp$ represents the number of particles per unit volume lying in the range from $p$ to $p+dp$ at a given time $t$.
Since the particles are highly relativistic, we can express the energy of the particle $E \approx p c$ ($c$ being  the speed of light) and therefore, ${\cal N}(E, \tau) dE = {\cal N}(p, \tau)dp$.
Integrating Eq. (\ref{eq:CR_p}) over the solid angle yields
\begin{equation}\label{eq:CR_E}
  \frac{d{\cal N}}{d\tau}
  + \frac{\partial}{\partial E}\left[\left(
  - \frac{E}{3} \nabla_{\mu}u^{\mu}  + \dot{E}_l
  \right) {\cal N}\right] = -{\cal N}\nabla_{\mu}u^{\mu} 
\end{equation}
where the first term in square brackets accounts for energy losses from adiabatic expansion while the second term $\dot{E}_l = \av{\dot{p}}_{l}/p^2$ is the radiative loss term due to synchrotron and IC processes.

\subsection{Radiative Losses}
\label{ssec:radloss}
%
%

Energetic electrons loose energy by  \textit{synchrotron} emission in the presence of magnetic fields and by the \textit{inverse Compton} (IC) process up-scattering the  surrounding radiation field.
For the latter process we assume that the scattering in the relativistic particle rest frame is Thompson, so  that the cross section $\sigma_T$ is independent of the incident photon energy $E_{\rm ph}$.
The energy loss terms for electrons with isotropically distributed velocity vectors is therefore given by:
\begin{equation}\label{eq:syncIC}
  \dot{E}_{l} = -c_rE^2
\end{equation}
where
\begin{equation}\label{eq:cr}
  c_r = \frac{4}{3}\frac{\sigma_T c \beta^{2}}{m_\e^2c^4}
        \left[ U_{\rm B}(t) + U_{\rm rad}(E_{\rm ph}, t)\right],
\end{equation}
while $\beta$ is the velocity of the electrons (we assume $\beta = 1$ for highly relativistic electrons) and $m_\e$ is their mass.
The quantities $U_{\rm B} =  \frac{B^{2}}{8\pi} $  and $U_{\rm rad}$ are the magnetic and the radiation field energy densities, respectively. 
For the present work, we use the isotropic Cosmic Microwave Background (CMB) as the radiation source. Therefore, applying the black body approximation, 
we have $U_{\rm rad} = a_{\rm rad} T^4_{\CMB} = a_{\rm rad} T_0^4 (1 + z)^{4}$ where  $a_{\rm rad}$ is the radiation constant, $z$ is the red-shift and $T_0 = 2.728$\,K is the temperature of CMB at the present epoch.

\subsection{Numerical Implementation}
\label{sec:lp}
%
%

Eq. (\ref{eq:CR_E}) is solved using a particle approach where a large number of Lagrangian (or macro-) particles sample the distribution function in physical space.
A macro-particle represents an ensemble of actual particles (leptons or hadrons) that are very close in physical space but with a finite distribution in energy (or momentum) space. 
To each macro-particle we associate a time-dependent energy distribution function ${\cal N}_\p(E, \tau)$ quantized in discrete energy bins.

For numerical purposes, however, it is more convenient to rewrite Eq. (\ref{eq:CR_E}) by introducing the number density ratio $\chi_\p = {\cal N}_\p/n$ which represents the number of electrons normalized to the fluid number density.
Using the continuity equation, $\nabla_\mu(n u^\mu) = 0$, it is straightforward to show that $\chi_\p$ obeys to the following equation:
\begin{equation}\label{eq:CR_chi}
  \frac{d\chi_\p}{d\tau} + \pd{}{E}\left[
\left(- \frac{E}{3} \nabla_{\mu}u^{\mu}  + \dot{E}_l\right)\chi_\p\right] = 0\,.
\end{equation}
The solution of Eq. (\ref{eq:CR_chi}) is carried out separately into a transport step (during which we update the spatial coordinates of the particles) followed by a spectral evolution step (corresponding to the evolution of the particle energy distribution).
These two steps are now described.

\subsubsection{Transport Step.}
%
%

Since the distribution function is carried along with the fluid, the spatial part of Eq. (\ref{eq:CR_chi}) is solved by advancing the macro-particle coordinates $\vec{x}_{\rm p}$ through the ordinary differential equations:
\begin{equation}\label{eq:ptrans}
  \frac{d \vec{x}_\p}{d t} = \vec{v}(\vec{x}_\p) \,,
\end{equation}
where $\vec{v}$ represents the fluid velocity interpolated at the macro-particle position and the subscript ${\rm p}$ labels the particle.
Eq. (\ref{eq:ptrans}) is solved concurrently with the fluid equations given by 
\begin{equation}
  \pd{\vec{U}}{t} + \nabla\cdot\tens{F} = \vec{S}
\end{equation}
which are solved as usual by means of the standard Godunov methods already present in the PLUTO code \citep{plutocode:2007, Mignone:2012}.
In the equation above, $\vec{U}$ is an array of conservative variables, $\tens{F}$ is the flux tensor while $\vec{S}$ denotes the source terms.

The same time-marching scheme used for the fluid is also employed to update the particle position. 
For example, in a $2^{\rm nd}$ order Runge-Kutta scheme, a single time update consists of a predictor step followed by a corrector step:
\begin{enumerate}
  \item \emph{Predictor step}: particles and conservative fluid quantities
  are first evolved for a \emph{full} step according to
  \begin{equation}
  \left\{\begin{array}{lcl}
    \vec{x}_\p^{*} &=& 
    \DS \vec{x}_\p^n + \Delta t^n \vec{v}^{n}(\vec{x}_\p^{n})
    \\ \noalign{\medskip}
    \vec{U}^{*} &=& 
    \DS\vec{U}^n - \Delta t^n\left(\nabla\cdot\tens{F}\right)^{n} 
  \end{array}\right.
  \end{equation}
  where $\Delta t^n$ is the current level time step, $\vec{x}_\p^n$ 
  denotes the particle's position at time step $n$ and 
  $\vec{v}^{n}(\vec{x}_\p^{n})$ is the fluid velocity interpolated at the
  particle position (at the current time level).

  \item \emph{Corrector step}: using the fluid velocity field obtained 
  at the end of the predictor step, particles and fluid are advanced to the 
  next time level using a trapezoidal rule: 
  \begin{equation}
    \left\{\begin{array}{lcl}
    \vec{x}_\p^{n+1} &=&\DS  \vec{x}_\p^n + \frac{\Delta t}{2}
    \Big[\vec{v}^{n}(\vec{x}_\p^{n}) + \vec{v}^{*}(\vec{x}_\p^{*})\Big]
    \\ \noalign{\medskip} 
    \vec{U}^{n+1} &=&\DS \vec{U}^{n} - \frac{\Delta t}{2}
                  \Big[   \left(\nabla\cdot\tens{F}\right)^{n}  
                        + \left(\nabla\cdot\tens{F}\right)^{*}\Big]
  \end{array}\right.
  \end{equation}
  where $\vec{v}^{*}(\vec{x}_\p^{*})$ denotes the (predicted) fluid 
  velocity interpolated at the (predicted) particle position.
\end{enumerate}

Interpolation of fluid quantities at the particle position is carried out by means of standard techniques used in Particle-In-Cell (PIC) codes 
\citep[see, e.g., the book by][]{Birsdall_and_Langdon.2004}
\begin{equation}
  \vec{v}(\vec{x}_\p) = \sum_{ijk} W(\vec{x}_{ijk} - \vec{x}_\p)\vec{v}_{ijk}
\end{equation}
where $W(\vec{x}_{ijk} - \vec{x}_\p) = W(x_i-x_\p)W(y_j-y_\p)W(z_k-z_\p)$ is the product of three one-dimensional weighting functions, while the indices $i$, $j$ and $k$ span the computational (fluid) grid.
For the present implementation we employ the standard second-order accurate triangular shaped cloud (TSC) method.

Particles are stored in memory as a doubly linked list in which each node is a C data-structure containing all of the particle attributes.   
The parallel implementation is based on the Message Passing Interface (MPI) and it employs standard domain decomposition based on the fluid grid.
Particles are therefore distributed according to their physical location and are thus owned by the processor hosting them.  
Parallel scaling up to $10^{4}$ processors has been demonstrated in a previous work, see \cite{Vaidya:2016}.

\subsubsection{Spectral Evolution Step.}
\label{ssec:ses}
%
%

As macro-particles are transported in space by the fluid, their spectral distribution evolves according to the energy part of Eq. (\ref{eq:CR_chi}) which can be regarded as a homogeneous scalar conservation law with variable coefficients in the $(E,\tau)$ space.
Here we show that a semi-analytical solution can be obtained using the method of characteristics.
The resulting expressions can then be used to advance the spectral energy distribution of the particles using a Lagrangian scheme in which the discrete energy grid points change in time.

To this purpose, we first observe that the characteristic curves of Eq. (\ref{eq:CR_chi}) are given by 
\begin{equation}\label{eq:CR_char_ode}
  \frac{dE}{d\tau} = -c_a(\tau) E - c_r(\tau)E^2 \equiv \dot{E}  \,,
\end{equation}
where $c_a(\tau) = \nabla_\mu u^\mu/3$, while $c_r(\tau)$ is given in Eq. (\ref{eq:cr}).
Integrating Eq. (\ref{eq:CR_char_ode}) for $\tau \ge \tau_0$, one finds 
\begin{equation}\label{eq:CR_char_sol}
  E(\tau) = \frac{E_0 e^{-a(\tau)}}{1 + b(\tau)E_0} \,,
\end{equation}
where $E_0$ is the initial energy coordinate while
\begin{equation}\label{eq:CR_a&b}
   a(\tau) = \int_{\tau_0}^\tau c_a(\tau)d\tau \,,\quad
   b(\tau) = \int_{\tau_0}^\tau c_r(\tau)e^{-a(\tau)} d\tau \,.
\end{equation}
Along the characteristic curve Eq. (\ref{eq:CR_chi}) becomes an ordinary differential equation so that, for each macro-particle, we solve
\begin{equation}\label{eq:CR_E2}
  \left.\frac{d\chi_\p}{d\tau}\right|_{\cal C}
  = - \left(\pd{\dot{E}}{E}\right)\chi_\p\,,
\end{equation}
where $\dot{E}$ is given by Eq. (\ref{eq:CR_char_ode}) while the suffix ${\cal C}$ on the left hand side denotes differentiation along the characteristic curve.
Integrating Eq. (\ref{eq:CR_E2}) and considering the fact that $\dot{E}$ is a function of $E$ alone, one finds
\begin{equation}\label{eq:chi_conserv}
  \chi_\p(E,\tau)dE = \chi_{\p0} dE_0\,,
\end{equation}
where $\chi_{\p0} = \chi_\p(E_0, \tau_0)$.
The previous expression shows that the number of particles (normalized to the fluid density) per energy interval remains constant as the interval changes in time.
The term $dE_0/dE$ describes the spreading or shrinking of the energy interval and it is readily computed from Eq. (\ref{eq:CR_char_sol}).
Integrating Eq. (\ref{eq:chi_conserv}) one has 
\begin{equation} \label{eq:chi_exact}
  \chi_\p(E(\tau), \tau) =
  \chi_{\p0} \Big[1 + b(\tau)E_0\Big]^2
              \left(\frac{n(\tau)}{n_0}\right)^{-1/3},
\end{equation}
where $n_0=n(\tau_0)$ and where we have used
\begin{equation}\label{eq:exp_a}
  e^{a(\tau)} = \exp\left(-\int_{\tau_0}^\tau \frac{d\log n}{3} \right)
              = \left(\frac{n(\tau)}{n_0}\right)^{-1/3}\,.
\end{equation}

The previous analytical expressions can be used to construct a numerical scheme based on a Lagrangian solution update.
To this purpose, we discretize (for each macro-particle) the energy space into $N_E$ energy bins of width $\Delta E^n_i = E^n_{i+\HALF}-E^n_{i-\HALF}$ (where $i=1,...,N_E$ while the superscript $n$ denotes the temporal index) spanning from $E^n_{\min}$ to $E^n_{\max}$. 
In our Lagrangian scheme, mesh interface coordinates are evolved in time according to Eq. (\ref{eq:CR_char_sol}) which we conveniently rewrite (using Eq. \ref{eq:exp_a}) as
\begin{equation}\label{eq:energy_interfaces}
  E^{n+1}_{i+\HALF} = \frac{E^n_{i+\HALF}}{1 + b^{n+1}E_{i+\HALF}^n}
                      \left(\frac{\rho^{n+1}}{\rho^n}\right)^{1/3} \,.
\end{equation}
The particle distribution $\chi_\p$ does not need to be updated explicitly (at least away from shocks, see section \ref{ssec:dsa}), since Eq. (\ref{eq:chi_conserv}) automatically ensures that the number of particle per energy interval is conserved time:
\begin{equation}
  \av{\chi}^{n+1}_{\p,i} = \frac{1}{\Delta E^{n+1}_i}
                          \int_{E^{n+1}_{i-\HALF}}^{E^{n+1}_{i+\HALF}}
  \chi_\p^{n+1}\,dE = \av{\chi}^{n}_{\p,i}  \,.
\end{equation}

This approach provides, at least formally, an exact solution update.
A numerical approximation must, however, be introduced since the coefficient $b(\tau)$ (Eq. \ref{eq:energy_interfaces}) has to be computed from fluid quantities at the particle position. 
Using a trapezoidal rule to evaluate the second integral in Eq. (\ref{eq:CR_a&b}) together with  Eq. (\ref{eq:exp_a}) we obtain
\begin{equation}\label{eq:b_new}
  b^{n+1} \approx \frac{\Delta t}{2}\left[
    \left(\frac{c_r}{\gamma}\right)^n
  + \left(\frac{c_r}{\gamma}\right)^{n+1}
     \left(\frac{\rho^{n+1}}{\rho^n}\right)^{1/3}\right] ,
\end{equation}
where the factor $1/\gamma$ comes from the definition of the proper time.
Eq. (\ref{eq:energy_interfaces}) with Eq. (\ref{eq:b_new}) do not make the scheme implicit inasmuch as the spectral evolution step is performed after the fluid corrector and the particle transport step.

Our method extends the approaches of, e.g.\cite{Kardashev:1962, Mimica:2012} and it is essentially a Lagrangian discretization for updating the distribution function in the energy coordinate.

In all of the tests presented here we initialize $\av{\chi}_\p$ at $t=t^0$ using an equally spaced logarithmic energy grid and a power-law distribution  
\begin{equation}\label{eq:pdistini}
  \av{\chi}^0_{\p,i} = \frac{{\cal N}_{\rm tot}}{n^0} \left(
   \frac{1 - m}{E_{\max}^{1-m} - {E_{\min}^{1-m}}}\right) E_i^{-m},
\end{equation}
where ${\cal N}_{\rm tot}$ is the initial number density of physical particles (i.e., electrons) associated to the macro-particle ${\rm p}$, $n^0$ is the initial fluid number density interpolated at the particle position and $m$ is the electron power index.

{\blue We remark that our formalism holds if physical microparticles embedded within a single macro-particle remain close in physical space (although they are allowed to have a distribution in energy space).
Therefore, an additional constraint should be imposed on the maximum Larmor radius so that it does not exceed the computational cell size.
This sets an upper threshold $E_{\rm max}$ to the maximum attainable energy of a given macroparticle, namely
\begin{equation}\label{eq:gamma_cf}
  E\le E_{\rm max} = \gamma_{L}^{cf}m_\e c^2  = \frac{e B r_{L}^{cf}}{\beta_{\perp}}
\end{equation}
where $r_{L}^{cf} = 0.5 \min(\Delta x, \Delta y, \Delta z)$ is computed at the particle cell position, $B$ is the magnetic field in Gauss and $e$ is electron charge in c.g.s. units. 
In the macro-particle limit, we assume that the individual leptons are highly relativistic and therefore the ratio of velocity of a single electron perpendicular to magnetic field with speed of light, $\beta_{\perp} \approx 1$.}

The initial energy bounds, the number of particles ${\cal N}_{\rm tot}$ as well as the value of $m$ are specified for each tests presented in this paper. The Lagrangian scheme described above has the distinct advantage of reducing the amount of numerical diffusion typical of Eulerian discretizations and it does not require explicit prescription of boundary conditions. 

\subsection{Diffusive Shock Acceleration}
\label{ssec:dsa}
%
%

The mechanism of diffusive shock acceleration (DSA) plays an important role in particle acceleration in a wide variety of astrophysical environments, particularly in Supernova remnants, AGN jets, GRBs, solar corona etc.  
The steady state theory of diffusive shock acceleration naturally results in power-law spectral distribution \citep[e.g.][]{Blandford:1978aa, Drury:1983aa, Kirk:2000aa, Achterberg:2001aa}
The two most important factors on which the post-shock particle distribution depends on are the strength of the magnetized shock (i.e. the compression ratio) and the orientation of magnetic field lines with respect to the shock normal.
The obliquity of magnetized shocks plays a very important role in determining the post-shock particle distribution \citep[e.g.,][]{Jokipii:1987, Ballard:1991}.
A comprehensive treatment was presented by \cite{Summerlin:2012} using Monte Carlo simulations, who have shown the importance of the  mean magnetic field orientation in the DSA process as well as the effect of MHD turbulence in determining the post-shock spectral index.
Analytical estimates of the spectral index for parallel relativistic shocks \citep{Kirk:2000aa, Keshet:2005aa} and for perpendicular shocks \citep{Takamoto:2015aa} have also shown remarkable consistency with the results from Monte Carlo simulations. 

\begin{figure}
  \centering
  \includegraphics[width=0.75\columnwidth]{\fpath/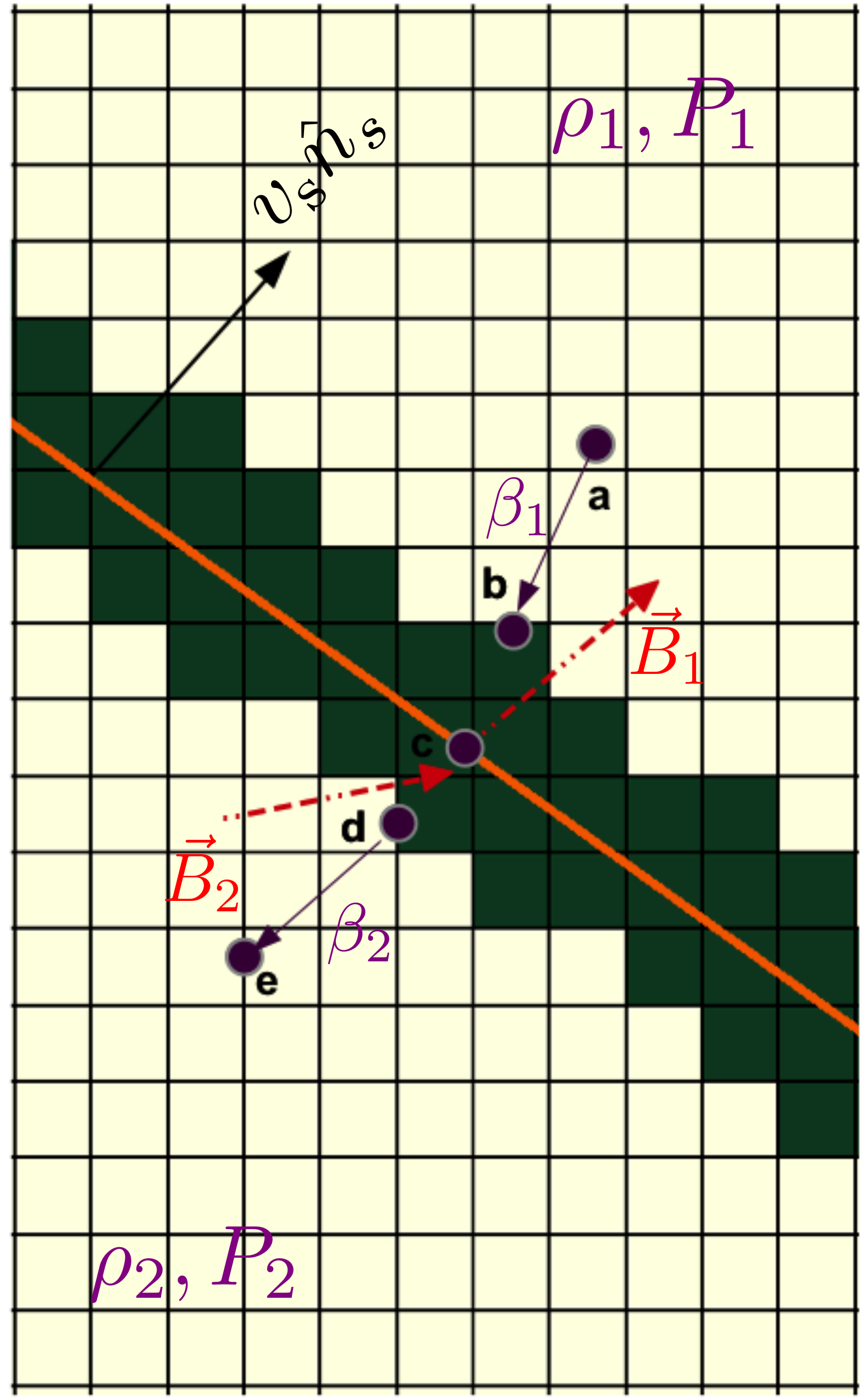}
  \caption{\footnotesize Cartoon figure showing the different positions of
           the particle and corresponding diagnostics.}
  \label{fig:shkp_cartoon}
\end{figure}

In our hybrid framework, modeling the post-shock spectral distribution with Monte Carlo method \citep{Summerlin:2012} is computationally very expensive and beyond the scope of present work. 
Instead we adopt the analytical estimates to account for DSA in the test particle limit valid for highly turbulent relativistic shocks. 
The slope of the spectral distribution associated with each macro-particle will depend on the compression ratio of the shock, $r$,  and the angle between the shock normal and magnetic field vector, $\Theta_{B}$. 
To estimate these quantities, we have devised a strategy based on a shock detection algorithm and the corresponding change in the energy distribution of the particle as it crosses the shock.
This is based on the following steps:
\begin{enumerate}
\item
We first flag computational zones lying inside a shock when the divergence of the fluid velocity is negative, i.e., $\nabla \cdot \vec{v} < 0$ and the gradient of thermal pressure is above a certain threshold, $\epsilon_{\rm sh}$ \citep[see also the appendix of][]{Mignone:2012}.
Typically we observe that a value  $\epsilon_{\rm sh} \sim 3$ is enough to detect strong shocks.
Shocked zones are shaded in green in Fig \ref{fig:shkp_cartoon}.

\item
Away from shocked zones (point $a$ in Fig \ref{fig:shkp_cartoon}), the particle spectral distribution evolves normally following the method outlined in the previous section.

\item
When the macro-particle enters a flagged zone (point $b$ in Fig \ref{fig:shkp_cartoon}), we start to keep track of the fluid state (such as density, velocity, magnetic field and pressure) by properly interpolating them at the macro-particle position.

\item
As the particle travels across the shocked area (points $b$, $c$ and $d$), we compute the maximum and minimum values of thermal pressure.
The pre-shock fluid state $\vec{U}_1$ is then chosen to correspond to the one with minimum pressure and, likewise, the post-shock state $\vec{U}_2$ to the one with maximum pressure. 
         
\item
As the macro-particle leaves the shock (point $d$), the pre- and post-shock states $\vec{U}_1$ and $\vec{U}_2$ are used to compute the orientation of the shock normal $\hvec{n}_s$ and thereafter the shock speed.
We employ the co-planarity theorem stating that the magnetic fields on both sides of shock front, $\vec{B}_{1}$ and $\vec{B}_{2}$, lie in the same plane as the shock normal, $\hvec{n}_{s}$.
Furthermore, the jumps in velocity and magnetic field across the shock must also be co-planar with the shock plane \citep{Schwartz:1998}.
By knowing two vectors co-planar to the plane of the  shock, we can easily obtain $\hvec{n}_{s}$ through their cross product:

 \begin{equation}
   \hvec{n}_s =  \begin{cases}
  \DS \pm \frac{\vec{\beta}^{\rm arb}_{2} - \vec{\beta}^{\rm arb}_{1}}{|\vec{\beta}^{\rm arb}_{2} - \vec{\beta}^{\rm arb}_{1}| } & \text{if  $\theta^{\rm arb} \approx 0^{\circ}$ or $90^{\circ}$}  
     \\ \noalign{\medskip}
  \DS \pm \frac{(\vec{B}_1 \times \Delta \vec{\beta}^{\rm arb}) \times \Delta \vec{B}}{|(\vec{B}_1 \times \Delta \vec{\beta}^{\rm arb}) \times \Delta \vec{B}|}& \text{otherwise }
\end{cases}
\end{equation}
where, $\vec{\beta}^{\rm arb}_{1,2}$ is the velocity vector in the pre- and post-shock states for an arbitrary frame (here the rest frame of underlying fluid) and $\theta^{\rm arb}$ is the angle between the magnetic field and the shock normal in that frame. The jumps in the fluid quantities are denoted $\Delta$, so that $\Delta \vec{B} = \vec{B}_{2} - \vec{B}_{1}$ and $\Delta \vec{\beta}^{\rm arb} = \vec{\beta}_{2} - \vec{\beta}_{1}$. Special care has to be taken to estimate shock normal in case of parallel and perpendicular shocks as the jump across the B field in the fluid rest frame will be zero (i.e., $\Delta \vec{B} = 0$)

We then compute the shock speed by imposing mass conservation of mass flux across the shock:
\begin{equation}\label{eq:mass_flux}
  \rho_{1} \gamma_{1} (\vec{\beta}_{1} - \vsh \hvec{n}_s)\cdot\hvec{n}_s  = 
  \rho_{2} \gamma_{2}(\vec{\beta}_{2} - \vsh \hvec{n}_s)\cdot \hvec{n}_{s}
\end{equation}
where the pre- and post-shock values are evaluated in the lab frame.
The previous equation holds also in the non-relativistic case by setting the Lorentz factors to unity.

\item
Next we compute the shock compression ratio $r$ defined as the ratio of upstream and downstream velocities in the shock rest frame ($\beta'_1$ and $\beta'_2$, respectively):
\begin{equation}
  r = \frac{\vec{\beta}_1^{\prime}
      \cdot \hvec{n}_s}{\vec{\beta}^{\prime}_2 \cdot \hvec{n}_s}
\end{equation}
In the the non-relativistic case, the shock rest frame can be trivially obtained  using a Galilean transformation.
In this case, the compression ratio can also be obtained from the ratio of densities across the shock (see Eq. \ref{eq:jumpmhd}).
However, this is no longer true in the case of relativistic shocks.
The reference frame transformation is not trivial in this case and multiple rest frames are possible.
In our approach, we transform from the lab frame to the \textit{Normal Incidence Frame}, NIF (see Appendix \ref{sec:frametrans}) to obtain the compression ratio using Eq. \ref{eq:jumprmhd}.

\item
The compression ratio $r$ and the orientation $\Theta_{B}$ of the magnetic field $\vec{B}$ with respect to the shock normal $\hvec{n}_{s}$ in the shock rest frame are used to update the particle distribution $\chi_{\rm p}(E, t^{d})$ in the post-shock region.
In particular, we inject a power-law spectrum in the post-shock region following ${\cal N}_\p(E, t^{d}) = {\cal N}(\epsilon_{0}) (E/\epsilon_0)^{-q + 2}$ where $\epsilon_0$ is the lower limit of the injected spectra and ${\cal N}(\epsilon_0)$ is the normalization constant.
These two quantities depends on two user-defined parameters viz., the ratio of non-thermal to thermal (real) particle densities, $\delta_n$,  and the ratio of total energy of the injected real particles to the fluid internal energy density $\delta_e$ \citep[see e.g.,][]{Mimica:2009, Boettcher:2010, Fromm:2016}.
In our recipe, we reset the particle energy distribution to the predicted DSA power-law by also taking into account the pre-existing population.
\color{black}
Therefore, we solve 

\begin{equation}\label{eq:Shk_Spec_1}
  \tilde{\cal A}(\epsilon_0) \int_{\gamma_0}^{\gamma_1}
  \left(\frac{\gamma}{\gamma_0}\right)^{-q + 2}d\gamma = \delta_n \frac{\rho}{m_{\rm i}} + n^{\rm old}
\end{equation}
and 
\begin{equation}\label{eq:Shk_Spec_2}
   \tilde{\cal A}(\epsilon_0) \int_{\gamma_0}^{\gamma_1}
    \left(\frac{\gamma}{\gamma_0}\right)^{-q + 2} \gamma d\gamma
   = \frac{\delta_e \mathcal{E}}{m_{\e} c^2} + \frac{E^{\rm old}}{m_{\e} c^2}
\end{equation}

to obtain the value of $\tilde{\cal A}(\epsilon_0)$ and $\epsilon_0 = \gamma_{0}m_{\e} c^2$. The number density $n^{\rm old}$  and energy $E^{\rm old}$ are obtained by integrating the spectra of the macro-particle before it has entered the shock.
In Eq. (\ref{eq:Shk_Spec_1}), $\rho$ is the value of fluid density interpolated at the macro-particle's position and $m_{\rm i}$ is the ion mass (we assume the thermal fluid density is dominated by protons).
${\cal E}$ is the fluid internal energy density interpolated at the particle position. 
Finally, the high energy cut-off, $\epsilon_1$ is estimated using the balance of synchrotron time scale, $\tau_{sync}$, to the acceleration time scale $\tau_{acc}$ \citep{Boettcher:2010, Mimica:2012}.
\begin{equation}\label{eq:Shk_Spec_3}
  \gamma_1 = \frac{\epsilon_1}{m_{\e}c^2} =
  \left(\frac{9 c^4 m_e^2}{8 \pi B \lambda_{\rm eff} e^3}\right)^{1/2} 
\end{equation}
where $m_e$ is the electron mass {\blue while the acceleration efficiency $\lambda_{\rm eff}$ is derived from the acceleration time scale as given by \citep{Takamoto:2015aa}
\begin{eqnarray}
\lambda_{\rm eff} &=& \frac{\eta r}{\beta_{1}^{\prime 2} (r - 1)}\left[\cos^2\Theta_{B 1} + \frac{\sin^2 \Theta_{B 1}}{1 + \eta^2} \right. \nonumber \\
 &&	\left. + \frac{rB_1^{\prime}}{B_2^{\prime}}\left(\cos^2 \Theta_{B 2} + \frac{\sin^2 \Theta_{B 2}}{1 + \eta^2}\right)\right]
\end{eqnarray}

where the dimensionless free parameter $\eta > 1$ is the ratio of gyro-frequency to scattering frequency and chosen to be a constant.
Primed quantities are computed in the shock rest frame.
The angles $\Theta_{B1}$ and $\Theta_{B2}$ represent the angles between the magnetic field vector and the shock normal in the upstream and downstream region.
We treat shocks to be quasi-parallel when $\eta \cos \Theta_{B 2} \ge 1$ and quasi-perpendicular otherwise (see \cite{Sironi:2009}). 
}

{\blue
\item The upper energy cutoff may exceed the maximum allowed energy imposed by the Larmor radius constraint (see Eq. \ref{eq:gamma_cf}).
If this is the case, we reset $\gamma_1$ to $\gamma_L^{cf}$ in order to avoid spatial spreading of microparticles.

\item As the macro-particle approach developed here aims at studying the large observable scales, micro-physical aspects of the DSA including amplification of magnetic fields, turbulent scattering have to be treated at the sub-grid level.
The free (dimensionless) parameter $\eta$ encapsulates the micro-physical nature of electron scattering associated with shock acceleration.
Studies of quasi-perpendicular relativistic shocks have shown small angle scattering as a dominant mechanism \citep{Kirk:2010, Sironi:2013aa} for accelerating electrons particularly in relativistic shocks.
In this regime, $\eta$ is a function of energy: $\eta \propto E$.
Here, for simplicity, we adopt a constant value of $\eta$ for both the downstream and the upstream flows.
This may over-estimate the acceleration efficiency particularly in the quasi-perpendicular case (see Eq.\ref{eq:Shk_Spec_3}).}

\item
The power-law index, $q$ for non-relativistic shocks used in our model is that obtained from steady state theory of DSA \citep{Drury:1983aa},
\begin{equation} \label{eq:qnonrel}
  q = q_{{\textsc{\tiny NR}}} = \frac{3 r}{r - 1}
\end{equation}
In case of relativistic shocks, power law index $q$ is obtained using analytical estimates from \cite{Keshet:2005aa} particularly under the assumption of isotropic diffusion, 
\begin{equation}\label{eq:qiso}
  q = \frac{3 \beta_{1}^{\prime} - 2 \beta_{1}^{\prime} \beta^{\prime 2}_{2}
            + \beta_{2}^{\prime 3}}{\beta_{1}^{\prime} - \beta_{2}^{\prime}}
    = q_{{\textsc{\tiny NR}}} + \left(\frac{1 - 2r}{r - 1} \right)\beta_{2}^{\prime 2}
\end{equation}
where $\beta_{1}^{\prime}$ and $\beta_{2}^{\prime}$ are the upstream and downstream velocity components along the shock normal in the NIF. 
In our test-particle framework, we assume isotropic diffusion for values of $\eta\cos\Theta_{B 2} \geq 1$ and use the spectral index from Eq. (\ref{eq:qiso}).
While for more oblique shocks we adopt the analytic estimate obtained by \citep{Takamoto:2015aa} for perpendicular shocks,
\begin{equation}\label{eq:qperp}
  q = q_{{\textsc{\tiny NR}}}
      + \frac{9}{20}\frac{r + 1}{r (r - 1)} \eta^{2}\beta_{1}^{\prime 2} 
\end{equation}

\item
Once the particle has left the shock (point $e$ in Fig \ref{fig:shkp_cartoon}, the distribution function is again updated regularly as explained in section \ref{ssec:ses}.

\end{enumerate}

\section{Emission and Polarisation Signatures}
\label{ssec:postproc}
%
%
In the previous sections, we described the framework and the methods used for following the temporal evolution of the distribution function of the ensemble of NTP attached to each Lagrangian macro-particle.  
The knowledge of the distribution function allows to compute the non-thermal radiation emitted by each macro-particle and from the spatial distribution of macro-particles we can reconstruct the spatial distribution of non-thermal radiation.
The non-thermal processes that we will consider are  synchrotron and IC  emission on a given radiation field and we will then be able to obtain intensity and polarization maps for each temporal snapshot. 
In the next subsection we describe synchrotron emission, while  subsection \ref{IC} will be devoted to IC radiation.

\subsection{Synchrotron Emission}
%
%

The synchrotron emissivity, in the direction $\hvec{n}^\prime$, per unit frequency and unit solid angle, by an ensemble of ultra-relativistic particles  is given by \citep[see][]{Ginzburg:1965}:
\begin{equation} \label{eq:emis}
  J_{\rm syn}^\prime(\nu^\prime, \hvec{n}^\prime)
  = \int \mathcal{P}^\prime(\nu^\prime, E^\prime, \psi^\prime)
          N^\prime(E^\prime, \hvec{\tau}^\prime) dE^\prime d\Omega^\prime_\tau
\end{equation}
where all primed quantities are evaluated in the local co-moving frame, which has a velocity $\vec{\beta} = \vec{v}/c$ with respect to the observer.
Here,  $\mathcal{P}^\prime(\nu^\prime, E^\prime, \psi^\prime)$ is the spectral power per unit frequency and unit solid angle emitted by a single ultra-relativistic particle, with energy $E^\prime$ and whose velocity  makes an angle $\psi^\prime$ with the direction $\hvec{n}^\prime$, while $N^\prime (E^\prime, \hvec{\tau}^\prime) dE^\prime d\Omega_\tau^\prime$  represents the number of particles with energy between $E^\prime$ and $E^\prime + dE^\prime$ and whose velocity is inside the solid angle $d\Omega^\prime_\tau$ around the direction $\hvec{\tau}^\prime$.
In performing the integrals, we can take into account that the particle radiative power, in the ultra-relativistic regime, is strongly  concentrated around the particle velocity and therefore only the particles with velocity along $\hvec{n}^\prime$ contribute to the integral, we can then set $N^\prime(E^\prime, \hvec{\tau}^\prime) = N^\prime(E^\prime, \hvec{n}^\prime)$.
Inserting in Eq. (\ref{eq:emis}) the expression for $\mathcal P$, that can be found in \citet{Ginzburg:1965}, we then get
\begin{equation} \label{eq:emis1}
J_{\rm syn}^{\prime}(\nu^\prime, \hvec{n}^\prime_{los}, \vec{B}^\prime)
  = \frac{\sqrt{3} e^3 } { 4 \pi m_e c^2}
    | \vec{B}^\prime \times \hvec{n}^{\prime}_{los} |
     \int_{E_i}^{E_f} \mathcal{N}^\prime (E^\prime) F(x) dE^\prime
\end{equation} 
where the direction individuated by $\hvec{n}^\prime_{los}$ is the direction of the line of sight,  we assumed that the radiating particles are electrons and we took a particle distribution that is isotropic and  covers an energy range between a minimum energy $E_{i}$ and a maximum energy $E_f$.
From the isotropic condition we can also write
\begin{equation}
  \mathcal{N}^\prime (E^\prime) = 4 \pi N^\prime(E^\prime, \hvec{n}^\prime). 
\end{equation}
Finally, the function $F(x)$ is the usual \textit{Bessel} function integral given by
\begin{equation}
  F(x) = x \int_{x}^{\infty} K_{5/3}(z)dz
\end{equation}
where the variable  $x$ is
\begin{equation}
  x = \frac{\nu^{\prime}}{\nu^{\prime}_{cr}} 
    = \frac{4 \pi m_e^{3} c^{5}\nu^{\prime}}
           {3eE^{\prime 2} |\vec{B}^{\prime} \times \hvec{n}^{\prime}_{los}| } 
\end{equation}
and $\nu^{\prime}_{cr}$ is the critical frequency at which the function, $F(x)$ peaks.
Similarly, the linearly polarized emissivity is given by
\begin{equation}  \label{eq:emispol}
  J^{\prime}_{\rm pol}(\nu^\prime, \hvec{n}^\prime_{los}, \vec{B}^\prime)
   = \frac{\sqrt{3} e^3 } { 4 \pi m_e c^2}
   | \vec{B}^\prime \times \hvec{n}^{\prime}_{los} |
   \int_{E_i}^{E_f} \mathcal{N}^\prime (E^\prime) G(x) dE^\prime
\end{equation}
where, the \textit{Bessel} function $G(x) = x K_{2/3}(x)$. 

Eqs. (\ref{eq:emis1}) and (\ref{eq:emispol}) give the emissivities in the co-moving frame as functions of quantities measured in the same frame, we need however the emissivities in the observer frame as functions of quantities in the same frame, these can be obtained by applying the appropriated transformations:
\begin{equation} \label{eq:Jsy}
  J_{\rm syn}(\nu, \hvec{n}_{los}, \vec{B})
  = \mathcal{D}^2  J_{\rm syn}^{\prime}(\nu^\prime, \hvec{n}^\prime_{los}, \vec{B}^\prime),
\end{equation}
\begin{equation} \label{eq:Jpol}
  J_{\rm pol}(\nu, \hvec{n}_{los}, \vec{B})
  =  \mathcal{D}^2 J^{\prime}_{\rm pol}(\nu^\prime, \hvec{n}^\prime_{los}, \vec{B}^\prime) 
\end{equation}
where the \textit{Doppler factor} $\mathcal{D}$ is given by
\begin{equation}
\mathcal{D}(\vec{\beta}, \hvec{n}_{los}) = \frac{1}{\gamma(1 - \vec{\beta} \cdot \hvec{n}_{los})}, 
\end{equation}
$\gamma$ is the bulk Lorentz factor of the macro-particle, while $\nu^\prime$,  $\hvec{n}^\prime_{los}$ and $\vec{B}^\prime$ can be expressed as functions of $\nu$,   $\hvec{n}_{los}$ and $\vec{B}$  through the following expressions:
\begin{eqnarray}
\nu^\prime & = & \frac{1}{\mathcal{D}} \nu \\
\hvec{n}^{\prime}_{los} & = &  
  \mathcal{D}\left[\hvec{n}_{los} 
   + \left(\frac{\gamma^{2}}{\gamma + 1} \vec{\beta}\cdot\hvec{n}_{los} 
   - \gamma\right)\vec{\beta}\right] \label{eq:nprime}\\
  \vec{B}^{\prime} & = & 
   \frac{1}{\gamma} \left[\vec{B} 
       + \frac{\gamma^{2}}{\gamma + 1} (\vec{\beta} \cdot \vec{B}) \vec{\beta}\right]
\end{eqnarray}

Using Eqs. (\ref{eq:Jsy}) and (\ref{eq:Jpol}), we can get  for each macro-particle the associated total  and polarized emissivities, at any time.
The values are then \textit{deposited} from the macro-particle on to the grid cells so as to give grid distributions of total and polarized emissivities, $\mathcal{J}_{\rm syn} (\nu, \hvec{n}_{los}, \vec{r})$  and $\mathcal{J}_{\rm pol} (\nu, \hvec{n}_{los}, \vec{r})$, as functions of the position $\vec{r}$.

Specific intensity maps can now be obtained by integrating the synchrotron emissivity, $\mathcal{J}_{\rm syn}(\nu, \vec{r})$ along the line of sight, in the direction $\hvec{n}_{los}$,
\begin{equation}\label{eq:StkI}
  I_\nu(\nu, X, Y) = \int_{-\infty}^{\infty} \mathcal{J}_{\rm syn}(\nu, X, Y, Z)  dZ,
\end{equation}
where we introduced a Cartesian observer's frame where the axis $Z$ is taken along the line of sight and the axes $X$ and $Y$ are taken in  the plane of the sky.  The total intensity represents the first \textit{Stokes}  parameter.
To compute the other \textit{Stokes} parameters, $Q_\nu$ and $U_\nu$ (neglecting circular polarization), we need to estimate the polarization angle, $\chi$. Such an estimate would require to account for proper relativistic effects like position angle swings \citep[][]{Lyutikov:2003, DelZanna:2006}.
The two Stokes parameter in the plane of sky are given by \citep[see,][]{DelZanna:2006}
\begin{eqnarray}
Q_\nu (\nu, X, Y) = \int_{-\infty}^{\infty} \mathcal{J}_{pol}(\nu, X, Y, Z) \rm{cos}\,2\chi dZ \label{eq:StkQ}\\
U_\nu (\nu, X, Y) = \int_{-\infty}^{\infty} \mathcal{J}_{pol}(\nu, X, Y, Z) \rm{sin} 2\chi   dZ \label{eq:StkU}
\end{eqnarray}
where \citep[see][]{DelZanna:2006}:
\begin{equation}
\cos (2 \chi) = \frac{q_X^2-q_Y^2}{q_X^2+q_Y^2},  \qquad \sin (2 \chi) = -\frac{2 q_X q_Y}{q_X^2+q_Y^2}
\end{equation}
and
\begin{equation}
  q_X = (1-\beta_Z) B_X - \beta_X B_Z,  \qquad q_Y = (1-\beta_Z) B_Y - \beta_Y B_Z
\end{equation}
and the polarization degree is
\begin{equation}
  \Pi =\frac{\sqrt{Q_\nu^2 + U_\nu^2}}{I_\nu}.
\end{equation}

\subsection{Inverse Compton Emission}
\label{IC}
%
%
%
The other important emission mechanism that we consider is the Inverse Compton Effect due to the interaction of relativistic electrons with a given radiation field.  
In the present work,  we will focus on the IC emission on seed photons  due to the isotropic CMB radiation. 

The co-moving  IC photon emissivity $\dot{n}^{\prime}_{\tmop{IC}}(\nu^\prime, \hvec{n}^\prime) = j^{\prime}_{\tmop{IC}} / h \nu^\prime  $(number of photons  per frequency interval per unit solid angle around the direction $\vec{n}^\prime$) is given by
\begin{equation}\label{eq:ICemiss}
\begin{aligned}
 \dot{n}^{\prime}_{\tmop{IC}} (\nu^{\prime}, \hvec{n}^\prime)  = 
 \int^{\infty}_0 d \varepsilon'_{ph}  \int d \Omega'_{ph} \int d E^\prime \int d \Omega^\prime_\tau \\ 
  N(E^\prime, \vec{\tau}) 
 \ c (1 - \vec{\beta_e} \cdot \vec{l}^\prime)  n_{\tmop{ph}}^\prime (\varepsilon'_{ph}, \vec{l}^\prime) \sigma
  (\varepsilon'_{ph}, \vec{l}^\prime, \nu^\prime, \hvec{n}^\prime) 
\end{aligned}
\end{equation}
where $n_{\tmop{ph}}^\prime (\varepsilon'_{ph}, \vec{l}') $ and $N^\prime(E^\prime, \vec{\tau}$) are, respectively,  the spectral density distribution of the seed photons, in the co-moving frame, as a function of photon energy $\varepsilon'_{ph}$ and
photon direction $\vec{l}^\prime$ and the electron distribution as a function again of energy $E^\prime$ and direction $\vec{\tau}$.
The factor $c (1 - \vec{\beta_e} \cdot \vec{l}^\prime) $ arises from the differential velocity between the photon and the electron, and $\vec{\beta}_e$ is the scattering electron velocity vector in units of $c$.
The scattering cross-section, $\sigma$,  depends, in principle, on the directions and energies of incident and out-going photons. 

The seed photons are the CMB photons, then in the observer frame have a black-body distribution with energy density
\begin{equation} \label{eq:ucmb}
  u_{\CMB} =  4 \frac{\sigma_B}{c} \left[ T_{\CMB} (1+z) \right]^4
\end{equation}
where $\sigma_B$ is the Stefan-Boltzmann constant, $T_{\CMB} = 2.728$K is the CMB temperature and $z$ is the red-shift of the source we study.
We approximate the black-body distribution with a monochromatic distribution with energy equal to the peak energy of the black-body, $\varepsilon_{CMB} = k_B T_{CMB}$, where $k_B$ is the Boltzmann constant.
If the flow moves at relativistic bulk speed ($\gamma >> 1$), the seed photons in the co-moving frame are bunched in the direction opposite to the macro-particle  velocity. 
The photon spectral energy distribution can be written as
\begin{equation} \label{eq:ncmb}
  n_{\tmop{ph}}^\prime (\varepsilon'_{\tmop{ph}}, \vec{l}^\prime)
  = \frac{\gamma u_{\CMB}}{\varepsilon_{\CMB} }
    \delta(\vec{l}^\prime - \hvec{\beta})
    \delta(\varepsilon'_{\tmop{ph}}- \gamma \varepsilon_{\CMB}),
\end{equation}
where $\hvec{\beta}$ is the unit vector in the direction of the macro-particle velocity and $\delta$  represents the Dirac function.
The electron distribution is assumed to be isotropic $N^\prime(E^\prime, \vec{\tau}) = \mathcal{N}^\prime(E^\prime) / 4 \pi$

The scattered photons are beamed along the direction of the scattering electron so that  $\hvec{n}^\prime = \vec{\tau}$ and emerge after scattering with average final energy
\begin{equation}\label{eq:ICscatter}
  h \nu^\prime \approx \left(  \frac{E^\prime}{m_e c^2}\right)^2 \varepsilon'_{\tmop{ph}}  (1 + \vec{\tau}^\prime \cdot  \hvec{\beta}). 
\end{equation}

Using the Thomson cross section, which is justified when the incident photon energy, in the electron frame,  is much less than the electron rest mass energy,  i.e. assuming $ \sigma (\varepsilon'_{\tmop{ph}}, \vec{l}^\prime, \nu^\prime, \hvec{n}^\prime) = \sigma_T$,  inserting Eqs. (\ref{eq:ucmb}), (\ref{eq:ncmb}), and (\ref{eq:ICscatter}) in Eq. (\ref{eq:ICemiss}) and taking into account the appropriate Lorentz transformations, we can finally express  the IC emissivity in the observer frame for each macro-particle as
\begin{eqnarray} \label{eq:JIC}
  J_{\rm IC} (\nu, \hvec{n}_{los}) =
  \left(  \frac{\mathcal{D}^2 m_e c^2}{2 \pi k_B} \right)
           \sigma_B \sigma_T T_{\CMB}^3 (1+z)^3
   \\ \nonumber 
  \left(  \mathcal{D} \Lambda \chi \right)^{1/2} \mathcal{N}
  \left( \sqrt{\frac{\chi}{\mathcal{D} \Lambda}} \right),
\end{eqnarray}
where $\mathcal{D}$ is the Doppler factor,
\begin{equation}
  \Lambda = \frac{1 +  \hvec{n}_{los} \cdot \hvec{\beta}}{1 + \beta}
\end{equation}
and
\begin{equation}
  \chi = \frac{h \nu}{k_B T_{\CMB} (1+z)}.
\end{equation}

As we do for the synchrotron emissivity, we can \textit{deposit} the IC emissivity  on to the grid cells so as to give the grid distribution of $\mathcal{J}_{IC} (\nu, \hvec{n}_{los}, \vec{r})$ and finally we can obtain  specific intensity maps by integrating along the line of sight.

\section{Numerical Benchmarks}
\label{sec:numtests}
%
%
%

In this section we report a suite of numerical benchmarks aimed at validating the correctness of our numerical implementation.
\subsection{Classical Planar Shock}
\label{sec:classical_shock}
%
%

\begin{figure*}
 \centering
  \includegraphics[width=1.8\columnwidth]{\fpath/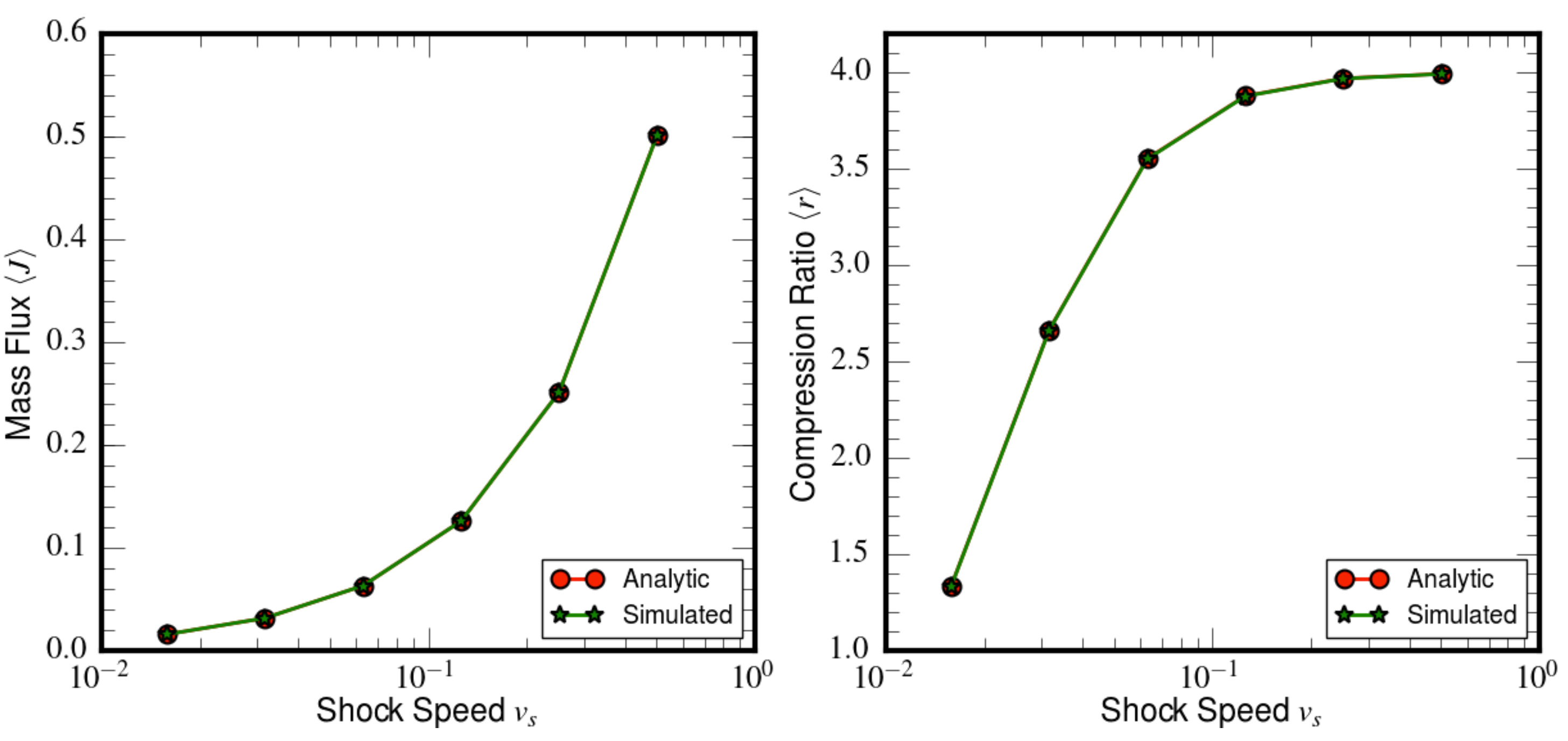}
  \figcaption{Analytical (\textit{red dots}) and simulated (\textit{green stars})
  values of the mass flux in shock rest frame $J$ (left panel) and compression
  ratio $r$ (right panel) for the classical MHD planar shock test with
  $\theta_B = 30^{\circ}$.} 
  \label{fig:cmpMHD}
\end{figure*}

In the first test problem we assess the accuracy of our method in verifying that the shock properties (such as compression ratio, mass flux, etc),  are sampled correctly as macro-particles cross the discontinuity.

We solve the classical MHD equation with an ideal equation of state ($\Gamma = 5/3$) on the Cartesian box $x\in[0,4]$, $y\in[0,2]$ using a uniform resolution of $512\times 256$ grid zones.
The initial condition consists of a planar shock wave initially located at $x_s(0) = 1$ and moving to the right with speed $\vsh$.
We work in the upstream reference frame where the gas is at rest with density and pressure equal to $\rho_1 = 1$, $p_1 = 10^{-4}$.
Here the magnetic field lies in the $x-y$ plane and it is given by $\vec{B} = B_1(\cos\theta_B,\, \sin\theta_B)$ where $\theta_B = 30^{\circ}$ is the angle formed by $\vec{B}$ and the $x$ axis while $B_1$ is computed from the plasma beta, $\beta_{p1} = 2p_1/B_1^2 = 10^2$.
The downstream state is computed by explicitly solving the MHD jump conditions once the upstream state and the shock speed $\vsh$ are known.
Zero-gradient boundary conditions are set on all sides.
We place a total of $N_p=16$ macro-particles in the region $1.5 < x < 3$ in the pre-shock medium and perform six different runs by varying the shock speed $\vsh \in[0.01, 1]$ on a logarithmic scale.

While crossing the shock, fluid quantities are interpolated at each macro-particle position following the guideline described in Sec \ref{ssec:dsa}.
From these values we compute, for each macro-particle, the mass flux $J_p$ and the compression ratio $r_p$ in the shock rest frame for each macro-particle. 
As all macro-particles experience the same shock, we compute the average value
\begin{equation}
  \av{J} = \frac{1}{N_\p}\sum_p J_\p
\end{equation}
and similarly for the compression ratio $\av{r}$.  
In the left and right panel of Fig. \ref{fig:cmpMHD} we compare, respectively, $\av{J}$ and $\av{r}$ with the analytical values obtained from the computations at different shock velocities.
Our results are in excellent agreement with the analytic values thereby demonstrating the accuracy of steps ${\rm i)}$ to ${\rm vi)}$ of the algorithm described in section \ref{ssec:dsa} in the non-relativistic case.

\subsection{Relativistic Planar Shock}
\label{sec:relativistic_shock}
%

\begin{figure*}
\centering
  \includegraphics[width=1.8\columnwidth]{\fpath/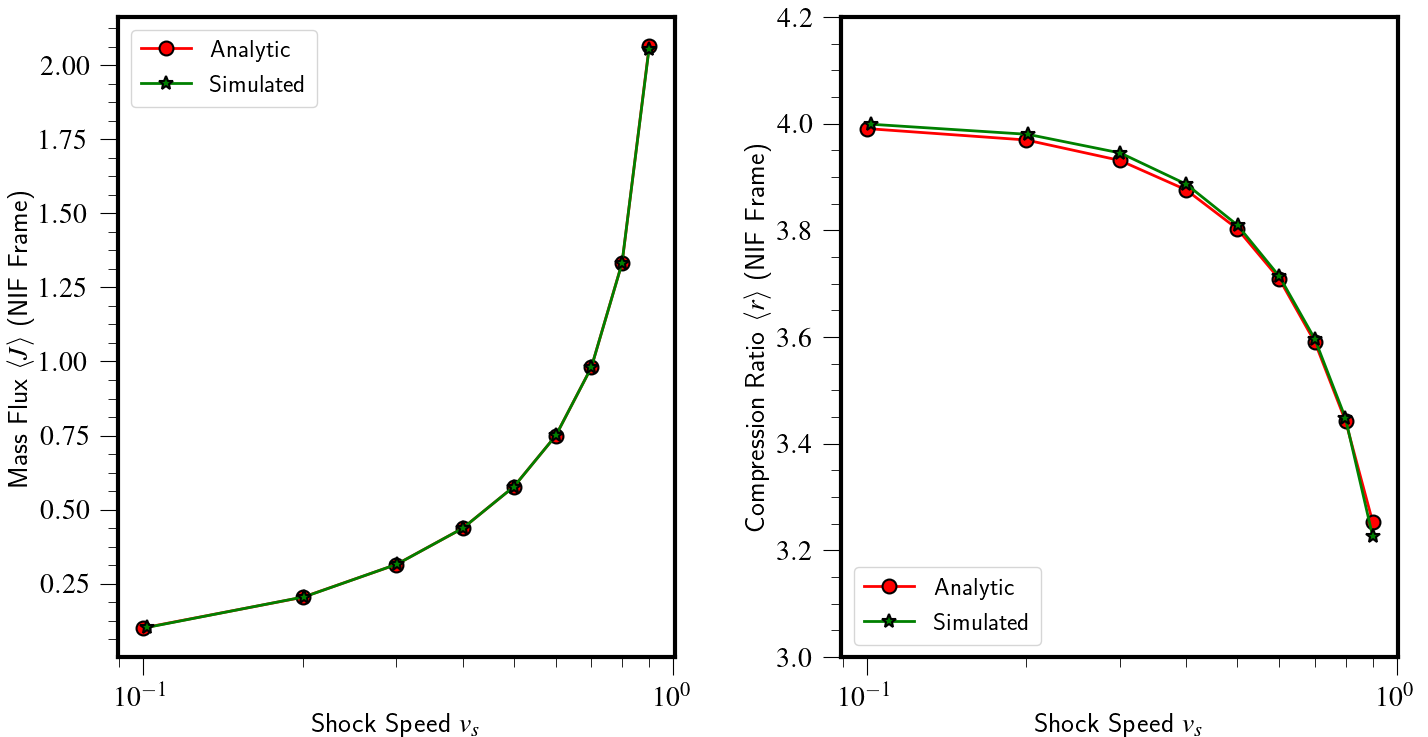}
  \figcaption{\textit{Left panel :} {\blue Comparison of mass flux and compression ratio for the relativistic planar shock case with $\Theta_B = 3.0$}. Analytical mass flux J in the lab frame (or NIF) estimated from Eq. (\ref{eq:mass_flux}) is shown as (\textit{red dots}), whereas its average value $\av{J}$ obtained from macro-particles are shown as \textit{green stars}. In the \textit{right panel}, the analytical (\textit{red dots}) 
  and simulated values of compression ratio, $r$ (\textit{green stars}) estimated using
  Eq. (\ref{eq:jumprmhd}) are shown.
  The simulated values are obtained as macro-particles traverse the relativistic planar shock and the
  sampled quantities across the shock are transformed to a shock rest frame.} 
\label{fig:cmpRMHD}
\end{figure*}

Next, we extend the previous problem to the relativistic regime with the aim to further describe the spectral evolution of macro-particles as they cross the discontinuity.
{\blue The initial conditions is similar to the previous test case but the upstream medium has now a transverse velocity $\vec{\beta} = 0.01 \hvec{y}$ and the magnetic field has a different strength given by $\beta_{p1} = 0.01$}.
We solve the relativistic MHD equation with the TM equation of state \citep{Taub:1948, Mathews:1971} and repeat the computation considering different values of the shock speed $\vsh$.
Like the classical case, we introduce $N_p=16$ macro-particles in the upstream reference frame in the region $1.5<x<3$.

As explained in section \ref{ssec:dsa}, we estimate relevant quantities such as the mass flux  $J$ and compression ratio $r$ by transforming to the Normal Incidence Frame (NIF) where the upstream velocity is normal to the shock front.
The strategy used for frame transformation is more involved than its classical counterpart and it is illustrated in Appendix \ref{sec:frametrans}. 

The left panel of Fig. \ref{fig:cmpRMHD} shows the analytical mass flux $J$ in the lab frame (see Eq. \ref{eq:mass_flux}) as \textit{red dots} and the average value of the mass flux $\av{J}$ obtained from the particles in the NIF frame as \textit{green stars}. A good agreement between the analytical and numerical results highlight the accuracy of our method in sampling the shock and the subsequent frame transformation required to quantify the compression ratio. 
A comparison between the analytical values (\textit{red dots}) for the compression ratio, $r$ with that obtained from macro-particles (\textit{green stars}) is shown in the right panel of Fig. \ref{fig:cmpRMHD}. 
We observe that the average compression ratio, $\av{r}$ estimated as the ratio of upstream and downstream velocities in NIF using macro-particles agrees with analytical values for varying shocks speeds.
The compression ratio value approaches, $r= 4.0$ for smaller shock speeds as expected from the non-relativistic limit.

\begin{figure*}
\centering
\includegraphics[width=1.7\columnwidth]{\fpath/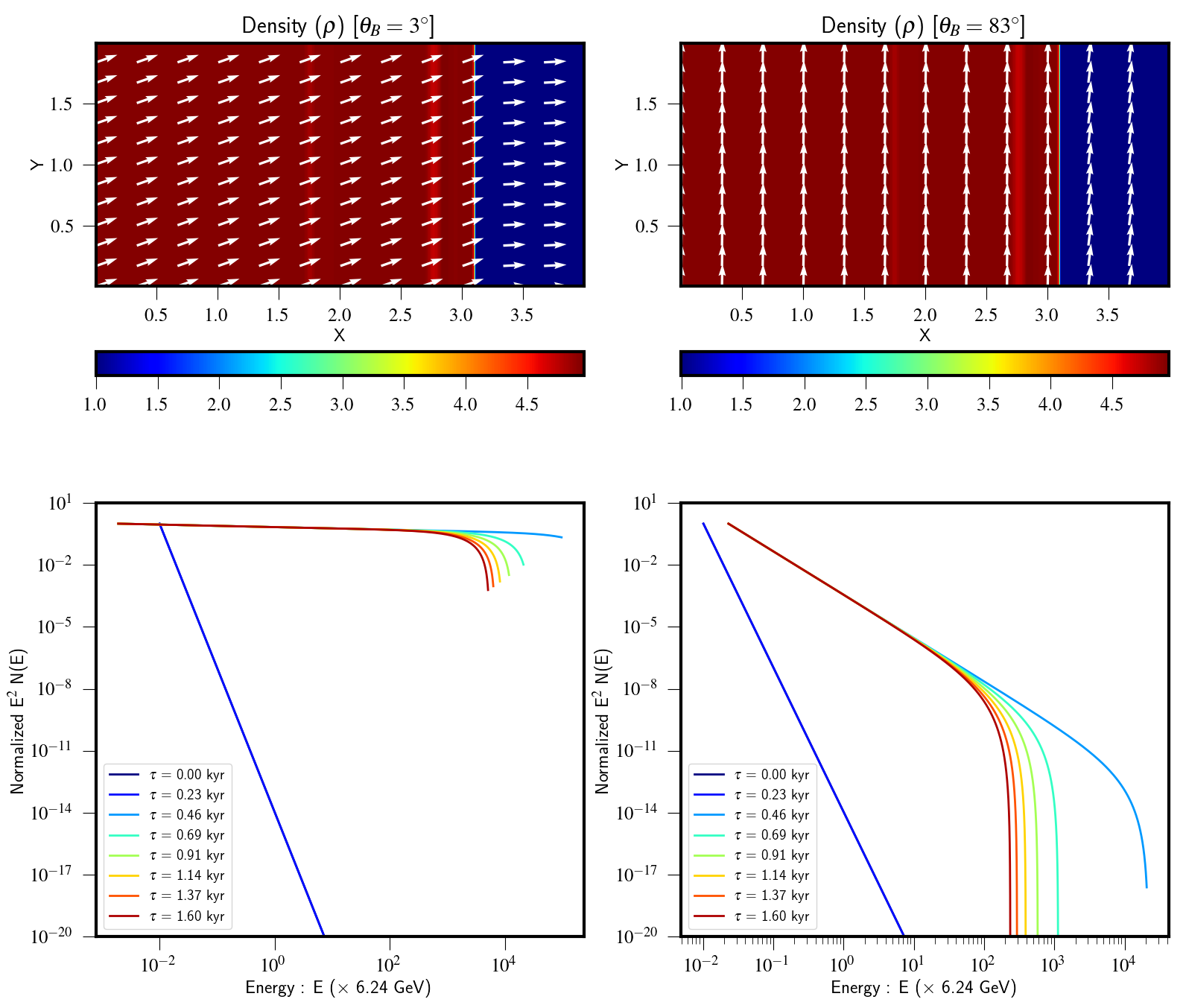}
\figcaption{\textit{Top panels} Density distribution in color at time $t = 0.98$ kyr along with magnetic field vectors shown as white arrows for quasi- parallel case  $\theta_B = 3^{\circ}$ (\textit{left})  and quasi-perpendicular case $\theta_B = 83^{\circ}$ (\textit{right}) for the relativistic planar shock test. \textit{Bottom panels} The corresponding evolution of normalized spectral distribution of a representative macro-particle. } 
\label{fig:cmpSpectraRMHD}
\end{figure*}

Next we focus on the evolution of the spectral energy distribution and, to this purpose, appropriate physical scales must be introduced.
We set the unit length scale $L_0 = 10^2\,{\rm pc}$ and the speed of light as the reference velocity, i.e. $V_0 = c$.
The energy distribution for each macro-particle is initialized as a power law with $m = 9$ (see Eq. \ref{eq:pdistini}) with the initial number density of real particles ${\cal N}_{\rm tot} = 10^{-4} \rm{cm}^{-3}$. 

The initial spectral energy ranges from $E_{\min} = 0.63$ MeV to $E_{\max} = 0.63$ TeV, with $n_E = 500$ bins. The initial bounds are chosen to cover a observed frequency range from radio band to X-rays for the chosen magnetic field strengths. The energy bounds of the spectral distribution as the macro-particles cross the shock are estimated from Eqns. (\ref{eq:Shk_Spec_1}) and (\ref{eq:Shk_Spec_2}) with $\delta_n = 0.9$ and $\delta_e = 0.5$.

We consider both quasi-parallel and quasi-perpendicular shocks where the angle between the shock normal and magnetic field vector is $\theta_B = 3^{\circ}$ and $\theta_B = 83^\circ$, respectively.
The shock jump conditions are set to obtain a compression ratio  $r \sim 3.6$ for both cases corresponding to the shock speed $v_s = 0.7 c$.
The density map and magnetic field orientation at  $t = 0.98\, {\rm kyr}$ are shown in the top panels of Fig \ref{fig:cmpSpectraRMHD} for the two cases.

The spectral evolution of a representative macro-particle are shown in the bottom panel of Fig. \ref{fig:cmpSpectraRMHD} for the quasi-parallel (left panel) and quasi-perpendicular (right panel) cases.
For the quasi-parallel case, the initial spectra steepens at high energies in presence of losses due to synchrotron emission.
{\blue At time $t \sim 0.46\, {\rm kyr}$ the macro-particle crosses the shock from the upstream region and the distribution function flattens its slope yielding a spectral index $q = 4.15$ as estimated from Eq. (\ref{eq:qiso}). 
Due to large acceleration time scale for quasi-parallel case, a high energy cutoff $E_{\rm max} \sim 6.25 \times 10^{5}\,{\rm GeV}$ is obtained as seen by the light blue curve in the left panel.
This sudden change in the spectra can be attributed to steady-state DSA, whereby the spectra is modified completely based on the compression ratio at the shock and particle density injected at shock (see Eqs.~\ref{eq:Shk_Spec_1} and ~\ref{eq:Shk_Spec_2}).
Subsequently, the high energy part cools down due to synchrotron emission reaching an energy of $\sim 10^{4}\,{\rm GeV}$ (red curve).
On the other hand, in the case of a  quasi-perpendicular shock, we obtain a steeper distribution owing to the dependence of the spectral index ($q = 6.2$) on $\eta^2$ (Eq. \ref{eq:qperp}).}
Also, the high energy cutoff lessens due to the inefficiency of quasi-perpendicular shocks in accelerating particles to high energy. 
The subsequent evolution of the particle spectrum is then governed by radiation losses due to synchrotron and inverse Compton cooling and lead to a similar steepening at high energies.
This test clearly shows the validity of our method in estimating the compression ratio $r$ and the change in the spectral slope under the DSA approximation.
\black
\subsection{Relativistic Magnetized Spherical Blast wave}
\label{sec:relativistic_blast}
%
%

In the next test case, we test our numerical approach on curved shock fronts to assess the accuracy of the method in the case where shock propagation is not grid-aligned. 

The initial conditions consists of a relativistic magnetized blast wave centered at the origin with density and pressure given by 
\begin{equation}
  (\rho,\, p) = \left\{\begin{array}{lll}
  (1,\, 1) & \quad\mathrm{for}\quad R < 0.8 l_0  \\ \noalign{\medskip}
  (10^{-2},\, 3\times 10^{-5})  & \quad\mathrm{otherwise}  &
  \end{array}\right.
\end{equation}
where $R = \sqrt{x^2+y^2}$ and $l_0$ is the scale length.
The magnetic field is perpendicular to the plane, $\vec{B} = B_0\hvec{z}$ with $B_0 = 10^{-6}$ and an ideal equation of state with adiabatic index $\Gamma = 5/3$ is used.

For symmetry reasons, we consider only one quadrant using $512^2$ computational zones on a square Cartesian domain of side $6 l_0$.
Reflecting conditions are applied at $x=y=0$ while outflow boundaries hold elsewhere.
The HLL Riemann solver, linear interpolation and a second-order Runge-Kutta are used to evolve the fluid.
We employ 360 macro-particles uniformly distributed between $0 < \phi < \pi/2$ and placed at the cylindrical radius $R_p = \sqrt{x_p^2 + y_p^2} = 2 l_0$.
Associated with each macro-particle is an initial power-law spectra with index $m = 9$ covering an energy range of 10 orders of magnitude with 500 logarithmically spaced uniform bins.

\begin{figure}
\centering
\includegraphics[width=1.\columnwidth]{\fpath/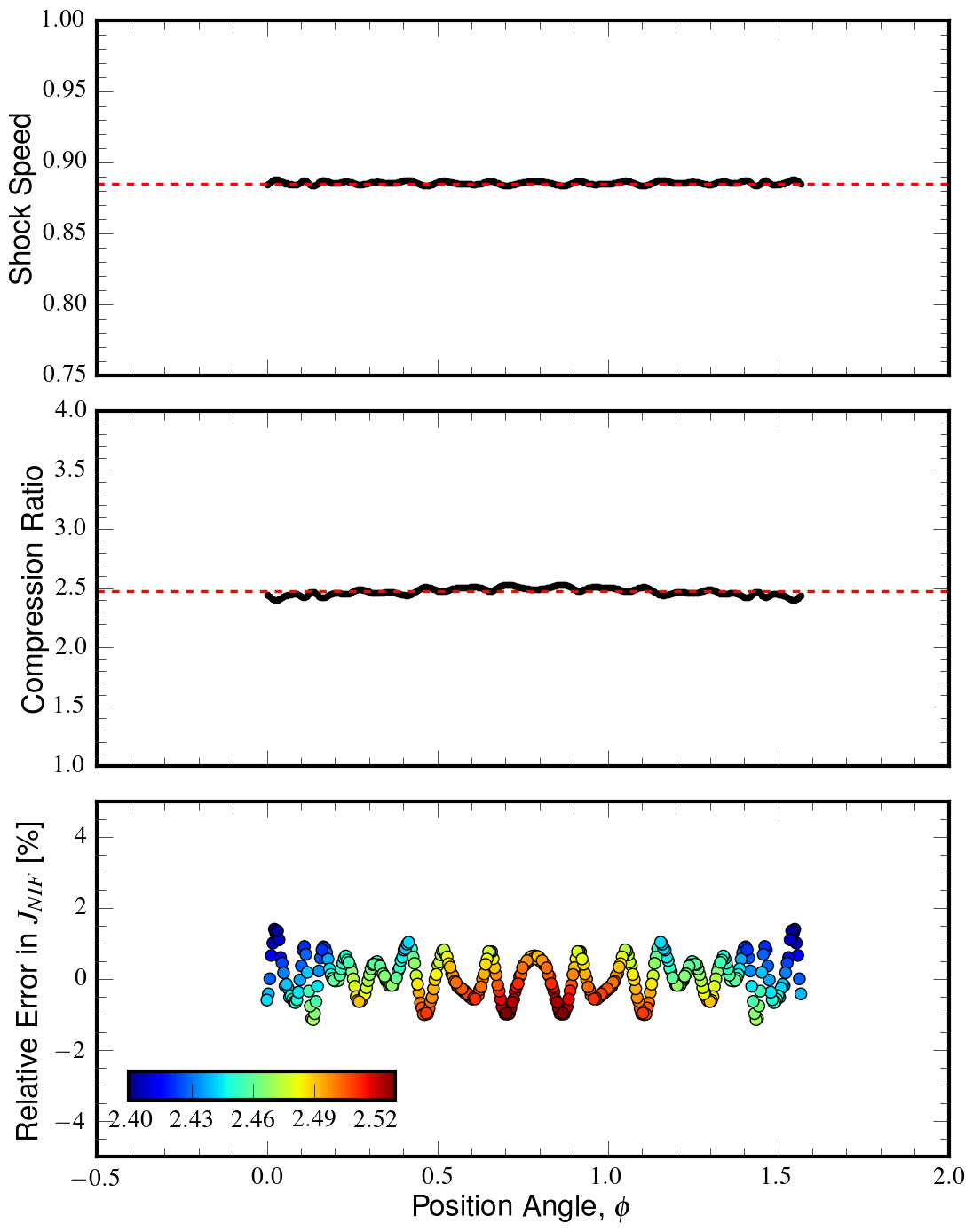}
\figcaption{The variation of shock properties with angular position for the RMHD blast wave test. The shock velocity obtained from a single representative macro-particle is shown as black circles and the semi-analytical estimate from a very high resolution 1D run is shown as a red dash line in the \textit{top panel}. The \textit{middle panel} shows the variation of compression ratio obtained from the particles. The relative error in the estimate of mass flux, $J_{\rm NIF}$ in the normal incidence frame is shown in the \textit{bottom panel}, the colors here indicate the value of compression ratio.} 
\label{fig:RMHDBlast2}
\end{figure}

The over-pressurized regions develops a forward moving cylindrical shock that propagates along the radial direction.
The shock velocity $\vsh$ computed by different macro-particles (see Sec. \ref{ssec:dsa}) is shown in the top panel of Fig \ref{fig:RMHDBlast2} as a function of the angular position and compared to a semi-analytical value $\vsh\approx 0.885$ obtained from a highly resolved 1D simulation. 
The numerical estimate of the shock speeds is consistent with the semi-analytical value within 1\% relative error. Additionally, its value remains the same independent of the angular position of the macro-particle. This clearly demonstrates the accuracy of our hybrid shock tracking method for curvilinear shock. 

This shock speed is then used to perform a Lorentz transformation to the NIF in order to obtain the compression ratio, shown in the middle panel of Fig.\ref{fig:RMHDBlast2}, from macro-particles initially lying at different angles. 
Similar to the shock velocity estimate, the compression ratio (middle panel) also agrees very well with the semi-analytical estimate $r \approx 2.473$ shown as a red dashed line.

The bottom panel of  Fig. \ref{fig:RMHDBlast2} shows the relative error in the estimate of mass flux, $J_{\rm NIF}$ in the normal incidence frame.
The relative error in the mass-flux is estimated as,
\begin{equation}
  \Delta J_{\rm NIF}[\%]  = 100 \left(\frac{J_{NIF}  - J_{NIF}^{ref}}{J_{NIF}^{ref}}\right) \,.
\end{equation}
where, the numerical mass flux down-stream of the shock in the NIF, $J_{NIF}$ is estimated from quantities interpolated on the macro-particles from the fluid. 
The reference value, $J_{NIF}^{ref}$, is estimated using the semi-analytical shock velocity and quantities across the shock from a highly resolved 1D simulation.
The color represents the value of the compression ratio as indicated from the color-bar.

\subsection{Sedov-Taylor Explosion}
%
%
\begin{figure*}
\centering
  \includegraphics[width=2.\columnwidth]{\fpath/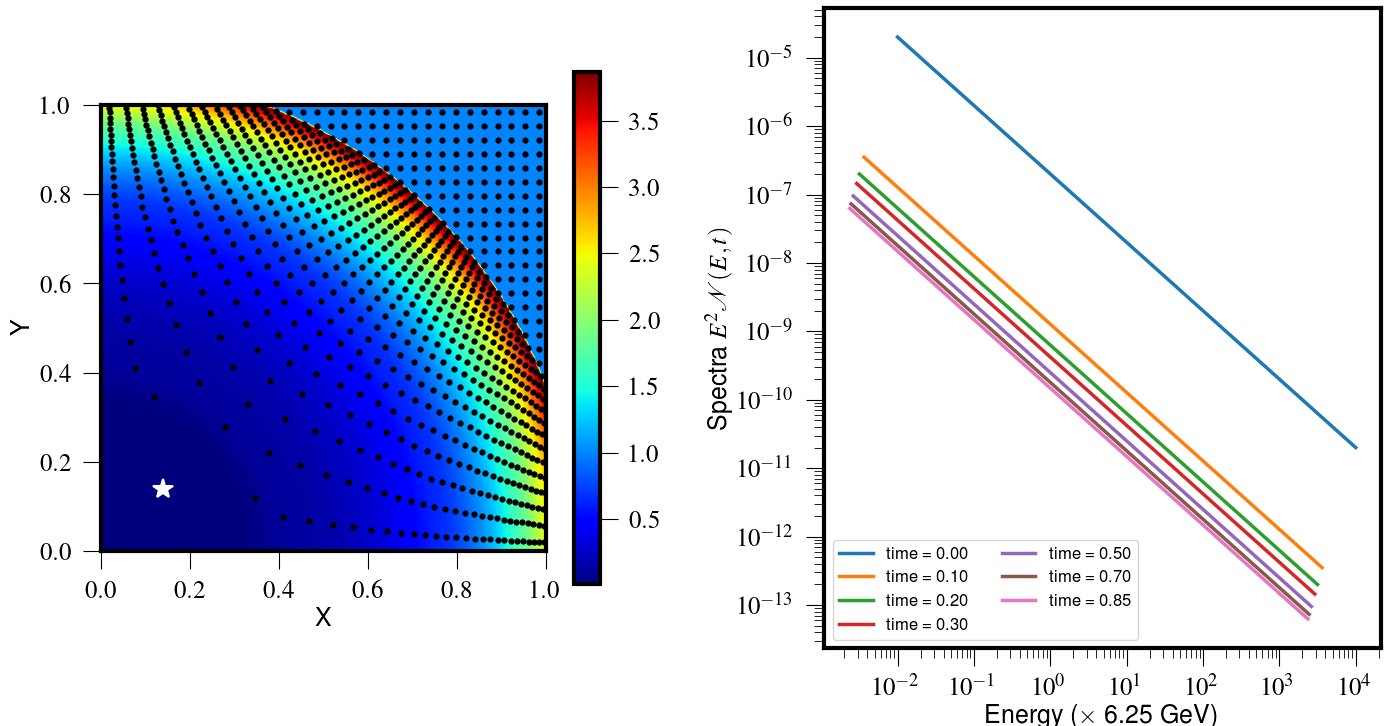}\\
  \figcaption{Left panel: particle distribution (black points) along with the
   fluid density (in color) for the Sedov Taylor explosion test at time $t = 0.85$ 
  (in arbitary units) with a resolution of $512^{2}$.
   Right panel: temporal spectral evolution for the macro-particle that
   is marked as white star in the left panel.}
\label{fig:sedov_pdist}
\end{figure*}

In the next test we verify the accuracy of our method in computing the radiative loss terms by focusing on the adiabatic expansion term alone, for which an analytical solution is available.
The fluid consists of a pure hydro-dynamical (B = 0) Sedov-Taylor explosion in 2D Cartesian coordinates ($x$,$y$) on the unit square $[0,1]$ discretized with $512^2$ grid points.
Density is initially constant $\rho=1$. A circular region around the origin ($x$=0, $y$=0) with an area $\Delta A = \pi \left(\Delta r\right)^2$ is initialized with a high internal energy (or pressure), where $\Delta r = 3.5/512$. While the region outside this circle has a lower internal energy (or pressure). Using an ideal equation of state with adiabatic index 5/3 we have, 
\begin{equation}
  \rho e = 
  \begin{cases}
    \DS  \frac{E}{\Delta A} &
     \text{for $r \le \Delta r$} \\ \noalign{\medskip}
      1.5  \times 10^{-5}   &  \text{otherwise}
   \end{cases}
 \end{equation}
where $r = \sqrt{x^2 + y^2}$ and input Energy, $E = 1.0$. Therefore we have contrast of 
$\approx 4.54 \times 10^8$ in $\rho e$.

For this test problem we have used the standard HLL Reimann solver with Courant number CFL = 0.4.
Reflective boundary conditions are set around the axis while open boundary conditions are imposed elsewhere.

Using the dimensional analysis, the \textit{self-similar} solution for the Sedov-Taylor blast can be derived.  
In terms of the scaled radial co-ordinate $\eta \equiv r (Et^2/\rho)^{-1/5}$, the shock location is obtained by: 
\begin{equation}
r_s(t) = \eta_s \left(\frac{Et^2}{\rho}\right)^{1/5} \propto t^{2/5}
\label{eq:ST_shkpos}
\end{equation}
where $\eta_s $ is a constant of the order of unity, {\blue $t$ is the time in \textit{arbitary} units} and $r$ is the spherical radius. The shock velocity follows via time differentiation as,
\begin{equation}
\vsh(t)  = \frac{d r_s}{d t} = \frac{2}{5} \frac{ r_{s} (t)}{t} \propto t^{-3/5}
\label{eq:ST_shkvel}
\end{equation}
Due to the self-similar nature, we can further relate the flow velocity at any spherical radius $r$ to that of the shock velocity obtained from Eq. \ref{eq:ST_shkvel}:
\begin{equation}
  v(r,t) \equiv \frac{\vsh(t)}{r_{s}(t)} r \equiv \frac{2}{5} r t^{-1}
\end{equation}
Thus, we have $\nabla \cdot \vec{v} \propto t^{-1}$. 
To estimate the evolution of spectral energy for a single macro-particle due to adiabatic expansion we have to solve Eq. (\ref{eq:CR_char_ode}) 
\begin{equation}\label{eq:ST_E(t)}
  E(t) = E_{0} {\exp}\left(-\int_{t_0}^{t} c_a(t)dt\right) 
\end{equation}
where $c_a(t) = \frac{1}{3}\nabla \cdot \vec{v}  \propto t^{-1}$.
Plugging Eq. (\ref{eq:ST_E(t)}) into Eq. (\ref{eq:chi_exact}) gives the temporal dependence of an initial power-law spectral density ${\cal N}(E, t)$:
\begin{equation} \label{eq:ST_N(t)}
{\cal N}(E, t)
  \propto E^{-m} \left(\frac{t_0}{t}\right)^{m + 2}
  \propto E_0^{-m} \left(\frac{t_0}{t}\right)^{2} \,,
\end{equation}
a result already known by \cite{Kardashev:1962}.

In order to compare the above analytical result with simulations, we initialize a total of 1024 macro-particles that are placed uniformly within the domain of unit square. 
Each particle is initialized with a power-law spectrum $\chi_\p^0 \propto E^{-m}$ (see Eq. \ref{eq:pdistini}) with $m=3$ covering a range of 6 orders of magnitude in the actual particle energy with a total of $250$ equally spaced logarithmic energy bins.
As the aim of this test is to study solely the effects due to adiabatic expansion, we switch off (by hand) the impact of shock acceleration due to the forward moving spherical shock.

Eq. (\ref{eq:ST_N(t)}) indicates that the ratio of spectral density varies with the inverse square law of time and does not affect the initial distribution slope $m$.
This implies that losses due to adiabatic expansion modify all energy bins in the same way and the resulting spectral evolution involves a parallel shift of the spectrum. Such an evolution of spectrum for a representative particle is shown in the right panel of Fig. \ref{fig:sedov_pdist}.

\begin{figure}
\centering
  \includegraphics[width=1.\columnwidth]{\fpath/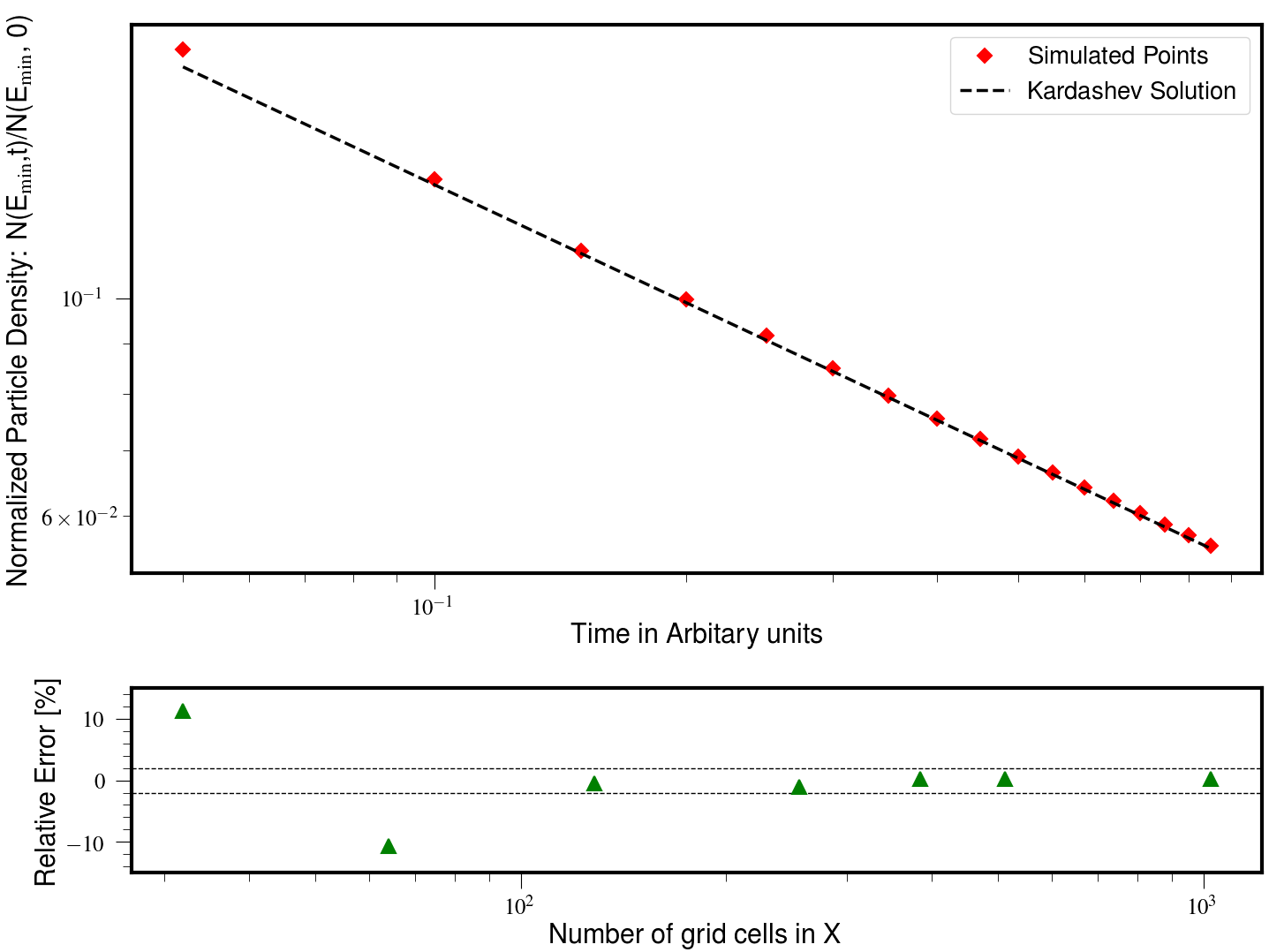}\\
  \figcaption{\textit{Top} Comparing the evolution {\blue with time (in arbitary units)} of normalized spectral distribution, $\mathcal{N}(E_{\rm min}, t)$ (red squares) with analytical solution obtained from Eq. \ref{eq:ST_N(t)} shown as black dashed line. \textit{Bottom} Results from the convergence study with different resolution are shown in this panel. The green triangles represent the relative errors (\%) in estimating the analytical slope for the variation of $\mathcal{N}(E_{\rm min}, t)$ with respect to time. The two black dashed line marks the $\pm$ 2\% error.}
\label{fig:sedov_cmpspectra}
\end{figure}

The particle distribution along with the fluid density (in color) at time $t = 0.85$ is shown in the left panel of Fig. \ref{fig:sedov_pdist}.
The particles that were initially placed uniformly have expanded with the flow as expected from their Lagrangian description.
Also, in the regions of high density just behind the shock, a large concentration of particles is seen.
The spectral evolution of the particle marked with \textit{white} color is shown in the right panel of the same figure.
Radiation losses due to adiabatic cooling affect all energy bins uniformly and, as a result, the spectra shifts towards the lower energy side keeping the same value of initial spectral power i.e., $m = 3$.
In order to test the accuracy of the numerical method applied, we have done a convergence study by varying the grid resolution of the unit square domain. 

In the top panel of Fig. \ref{fig:sedov_cmpspectra}, we compare the spectral distribution for a particular energy bin ($E = E_{\rm min}(t)$) of a single particle under consideration with the analytical solution described above.
We observe that for the  run with $512^{2}$ resolution, the simulated values are in perfect agreement with the analytical estimates.
However, the errors in the estimate of the slope becomes as large as 10-15\% with low resolution. 
The bottom panel of Fig. \ref{fig:sedov_cmpspectra} shows the relative error in \% for the estimate of the slope for different grid resolutions. The error is visibly large for grid resolutions $< 100$ points.
However, having more than 128 points in the domain results in reducing the error within the $\pm 2\%$ band as indicated by two black dashed lines and is fully converged for runs with 512 grid points.

\subsection{Relativistic Spherical Shell}
%
%
In this test, we verify our numerical implementation to estimate synchrotron emissivities (Eqns ~\ref{eq:Jsy} and ~\ref{eq:Jpol}) and the polarization degree from stokes parameters (Eqns ~\ref{eq:StkQ} and ~\ref{eq:StkU}) specifically testing the changes due to relativistic effects.

\subsubsection{Comoving frame}
%
%
We initialize a magnetized sphere in a three dimensional square domain of size $L = 40$\,pc. The sphere has a constant density ($\rho_0= 1.66 \times 10^{-25}$ g cm$^{-3}$) and pressure ($P_0 = 1.5 \times 10^{-4}$ dyne cm$^{-2}$ ) and is centered at the origin and has a radius of $R_s = 10$\,pc. The three components of the velocity are given such that,
\begin{equation}
\vec{v} =  \beta \frac{R}{R_s} \{ \sin(\theta)\cos(\phi), \sin(\theta)\cos(\phi), \cos(\theta)\}
\end{equation} 
 where $\beta = \sqrt{1 - 1/\gamma^2}$ with bulk Lorentz factor $\gamma$ and $R, \theta$ and $\phi$ are spherical co-ordinates expressed using Cartesian components.

The Cartesian components of the purely toroidal magnetic fields are set as follows,
\begin{eqnarray}
  B_{x} & = &  -B_0  \,\, \sin(\phi) \sqrt{x^2 + y^2}    \nonumber \\
  B_{y} & = &  B_0   \,\, \cos(\phi) \sqrt{x^2 + y^2}   \nonumber \\
  B_{z} & = &  0.0
\end{eqnarray} 
where $B_0 \sim 60$ mG is the magnitude of magnetic field vector.

A total of 100 macro-particles with an initial power-law spectral distribution are randomly placed on the shell of width $0.1 R_{s}$. For each particle, the spectral range from $E_{\min} = 10^{-8}$ ergs to $E_{\max} = 10^{2}$ ergs is sampled by a total of 250 logarithmically spaced energy bins.  The synchrotron emissivity, $J_{sy}(\nu)$ and linearly polarized emissivity $J_{pol}(\nu)$ from each of this macro-particle is estimated numerically using Eqs. \ref{eq:Jsy} and \ref{eq:Jpol} for an observed frequency $\nu = 10^{10}$\, GHz with the initial power-law spectral distribution. Their ratio gives a value of polarization fraction $\Pi_{i}$, for $i$th macro-particle. We compute the arithmetic average of numerically estimated polarization degree and is denoted by $\left<\Pi\right>$

\begin{figure}
\centering
\includegraphics[width=1.\columnwidth]{\fpath/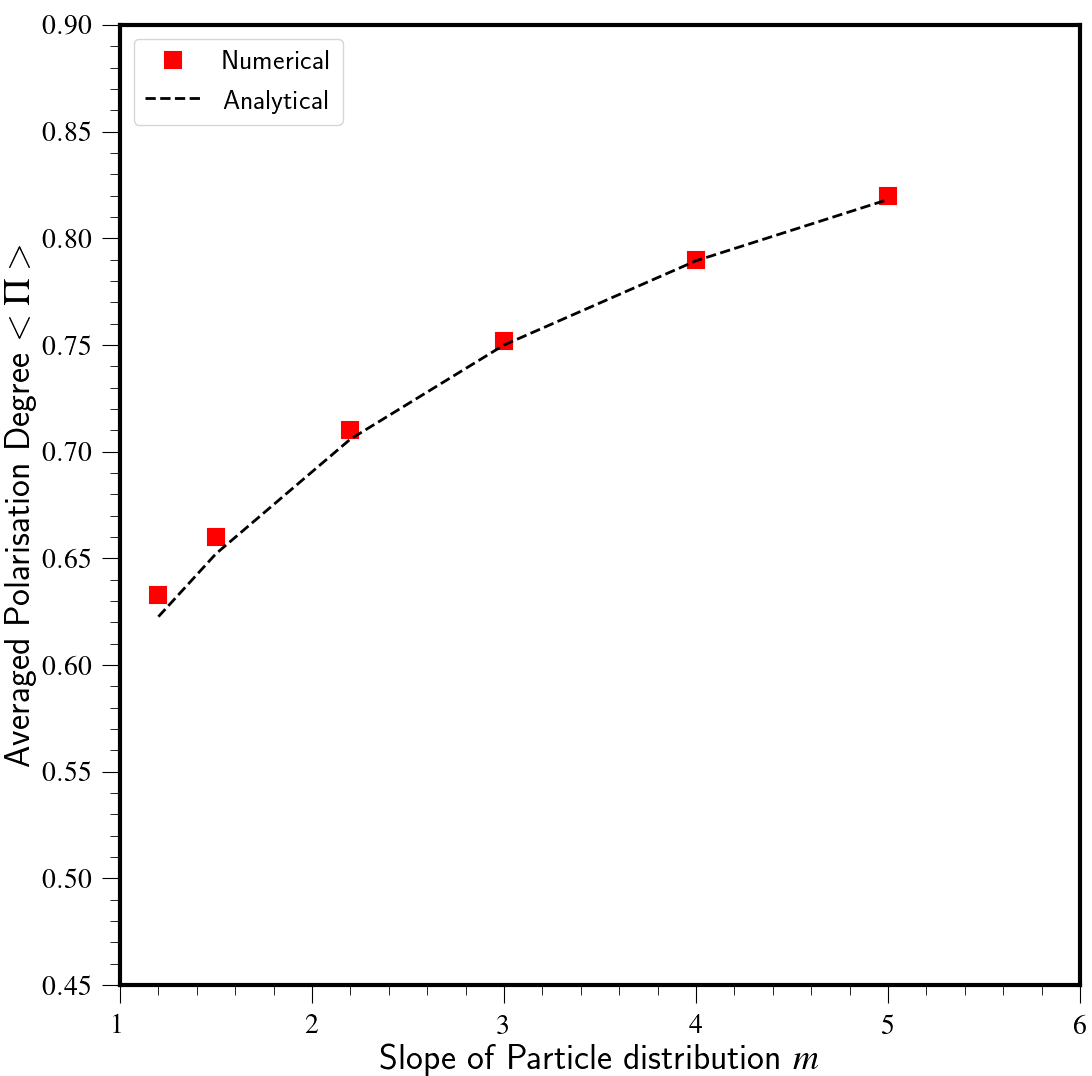}\\
\figcaption{Comparison of the numerically estimated averaged ratio of $J_{sy}(\nu)$ with $J_{pol}(\nu)$ for $\nu = 10^{10} GHz$ (red squares) with the theoretical values obtained from Eq.\ref{eq:poldeg_th} shown as black dashed line.}
\label{fig:poldeg}
\end{figure}

In the co-moving frame, the theoretical value expected for the polarization degree, on the shell is simply given by  \citep[e.g.][]{Longair:1994}
\begin{equation}  \label{eq:poldeg_th}
  \Pi = \frac{m + 1}{m + 7/3}.
\end{equation}
In figure Fig.~\ref{fig:poldeg}, we have compared the numerical averaged value (in co-moving frame) for different initial power-law spectral slope, $m$ with the above theoretical estimate (Eq. \ref{eq:poldeg_th})

\subsubsection{Observers Frame}

To obtain the polarization degree, $\Pi_{obs}$ in the observer frame, the Stokes parameters given by Eqns ~\ref{eq:StkQ} and ~\ref{eq:StkU} have to be computed along with the polarization angle $\chi$. Relativistic effects like position angle swing must be taken into account in order to calculate $\chi$ \citep[e.g.][]{Lyutikov:2003, DelZanna:2006}. Due to the  relativistic motion,  the emission is boosted, resulting in a rotation of linear polarization angle in the $\hvec{n}-\vec{v}$ plane. Though the value of fractional polarization is same, the rotation of polarization angle is different for different elements of the emitting object.  These relativistic kinematic effects can therefore result in the maximum observed polarization to be smaller than the theoretical upper limit given by Eq. ~\ref{eq:poldeg_th}.
This crucial ingredient has been implemented in our hybrid framework to compute the Stokes parameters and thereby the corrected fractional polarization in case of macro-particles moving in relativistic flow. 
Here, we verify our numerical implementation by replicating the calculation of the averaged value of the Stokes parameters done by \cite{Lyutikov:2003} for a quasi-spherical thin emitting shell. 

\begin{figure*}
\centering
\includegraphics[width=2.\columnwidth]{\fpath/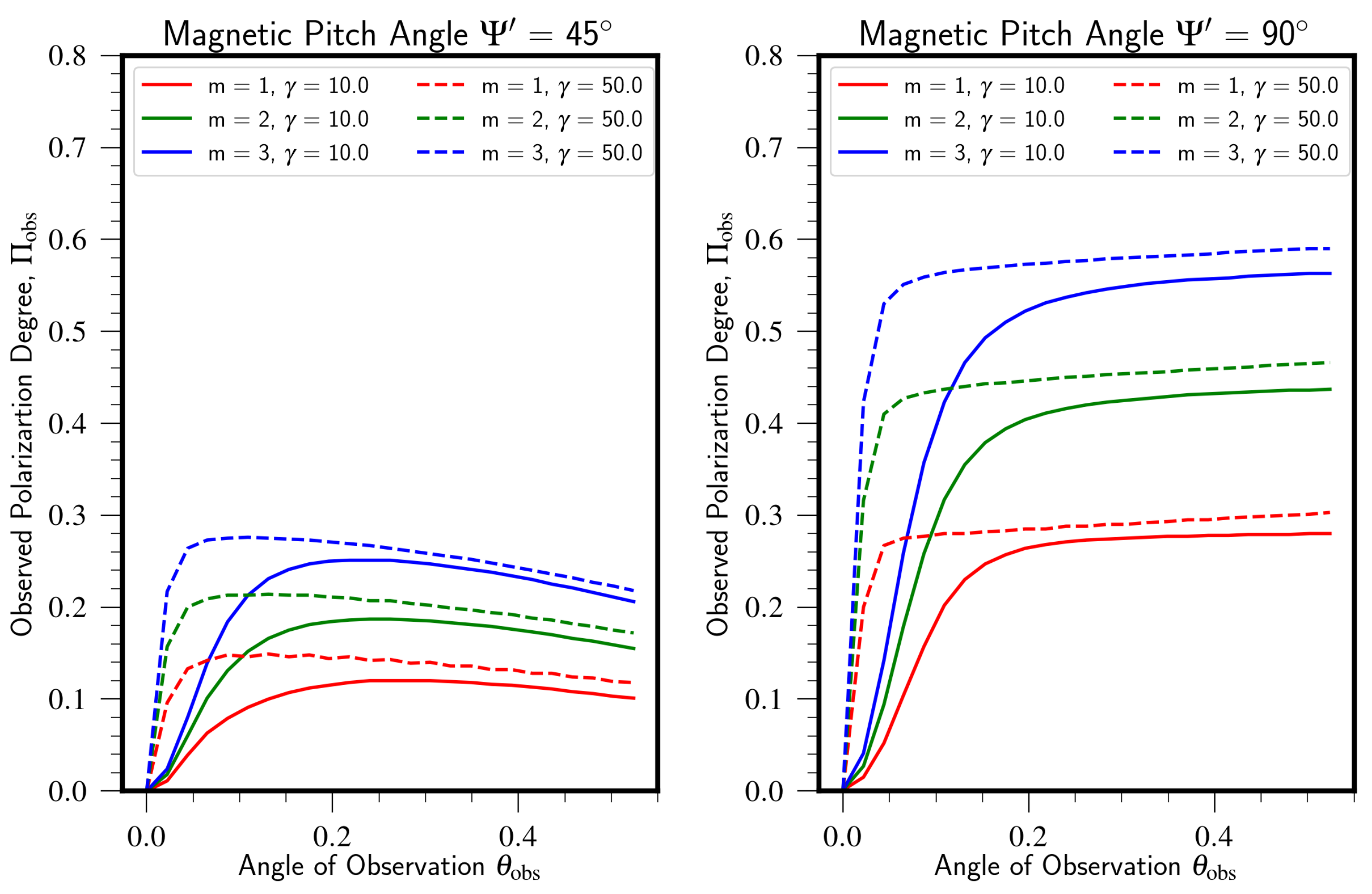}\\
\figcaption{\textit{Left: } Dependence of Observed polarization fraction, $\Pi_{\rm obs}$ with observation angle, $\theta_{\rm obs}$ for a magnetic pitch $\Psi^{\prime} = 45^{\circ}$ and two values of Lorentz factor for the shell, $\gamma$ = 10 (\textit{solid line}) and 50 (\textit{dashed line}). The macro-particles distribution (radiating elements) is set to be a power law with three different spectral slope, $m$ = 1(red), 2(green) and 3(blue). \textit{Right :} Same as the left panel but for a purely toroidal field ($\Psi^{\prime} = 90^{\circ}$).}
\label{fig:polgamma}
\end{figure*}

In our case, an emitting element is represented by a macro-particle that is moving with the spherical shell with a velocity that depends on the two  spherical co-ordinates $\theta$ and $\phi$:
\begin{equation}
  \vec{v} = \beta\{ \sin\theta \cos\phi, \sin \theta \cos \phi, \cos \theta\},
\end{equation}
where $\beta$ is a related to the Lorentz $\gamma$. The observer is  in the $x$-$z$ plane with 
\begin{equation}
  \hvec{n} = \{\sin \theta_{\rm obs}, 0, \cos \theta_{\rm obs}\}
\end{equation}
as the unit vector along the line of sight and $\theta_{\rm obs}$ is the angle with respect to vertical $z$-axis.
The shell is magnetized with a field that lies along -
\begin{equation}
\hvec{B} = \{-\sin \Psi^{\prime} \sin \phi, -\sin \Psi^{\prime} \cos \phi,  \cos \Psi^{\prime}\}
\end{equation}
where $\Psi^{\prime}$ is the magnetic pitch angle. Macro-particles that are placed uniformly on such a shell will emit synchrotron emission based on their initial power-law spectra govern by the index $m$ (same for all macro-particles).
The dependence of volume averaged stokes parameters obtained from our numerical implementation for two values of Lorentz $\gamma = 10$ (\textit{solid lines}) and $\gamma = 50$ (\textit{dashed lines}) of the shell and three values of initial power-law index (i.e., $m$ = 1, 2 and 3) of the emitting macro-particles is shown in Fig. ~\ref{fig:polgamma}. 
The left panel of the figure is for a value of the magnetic pitch angle $\Psi^{\prime} = 45^{\circ}$ and the right panel is for a purely toroidal field $\Psi^{\prime} = 90^{\circ}$.

For the case of a purely toroidal magnetic field, we observe that  the value of the  polarization degree saturates for $\theta_{\rm obs} > 1/\gamma$ consistent with the  electro-magnetic model proposed to explain large values of polarization reported in GRB \citep{Lyutikov:2003}.
As expected, the polarization fraction saturates at a smaller $\theta_{\rm obs}$ for $\gamma = 50$  as compared to runs with $\gamma = 10$.
The asymptotic value, $\Pi \approx 56\%$ obtained for $m = 3 (\textit{blue})$ is less than the maximum upper limit of 75\% (using Eq. ~\ref{eq:poldeg_th}),  in agreement with the estimates by \cite{Lyutikov:2003}. 
The effect of depolarization is further enhanced if the magnetic field distribution is changed using the value of $\Phi^{\prime} = 45^{\circ}$ (\textit{left panel}). 
In this case, the asymptotic value of  the polarization degree for $m = 3$ is $\leq 30\%$. This clearly shows the vital role of (de)-polarization degree in determining the magnetic field structure in the flow. 

\section{Astrophysical Application} 
\label{sec:application}
%
%
In this section, we describe  couple of astrophysical applications of the hybrid framework.

\subsection{Supernova Remnant SN1006}
\label{sec:snr1006}
%
%
The first application is to study classical DSA and properties of non-thermal emission from a historical Type IA Supernova remnant (SNR), SN1006.
The numerical setup chosen for this problem is identical to \cite{Schneiter:2010}. 
We perform axi-symmetric magneto-hydrodynamic simulation with a numerical grid of physical size of 12 and 24 pc in the $r$- and $z$-directions, respectively. 
The grid has a spatial resolution of $1.56\times10^{-2} $ pc. The ambient ISM has a constant number density, $n_{\rm amb} = 0.05\,{\rm cm}^{-3}$. 
The initial magnetic field is chosen to be constant with a value of 2\,$\mu$\,G and parallel to the $z$-axis.
To numerically model the Type Ia SNR, we initialize a sphere with radius of 0.65\,pc at the center of the domain such that it contains an ejecta mass of $1.4 M_{\odot}$. 
Within the sphere, the  innermost region has a constant mass equivalent to $0.8 M_{\odot}$ while the  rest of the mass is in the outer region. 
This outer region has an initial power-law density profile, $\rho \propto R_{\rm sph}^{-7}$, where the spherical radius $R_{\rm sph} = \sqrt{r^{2} + z^{2}}$.

\begin{figure}\centering
  \includegraphics[width=1\columnwidth]{\fpath/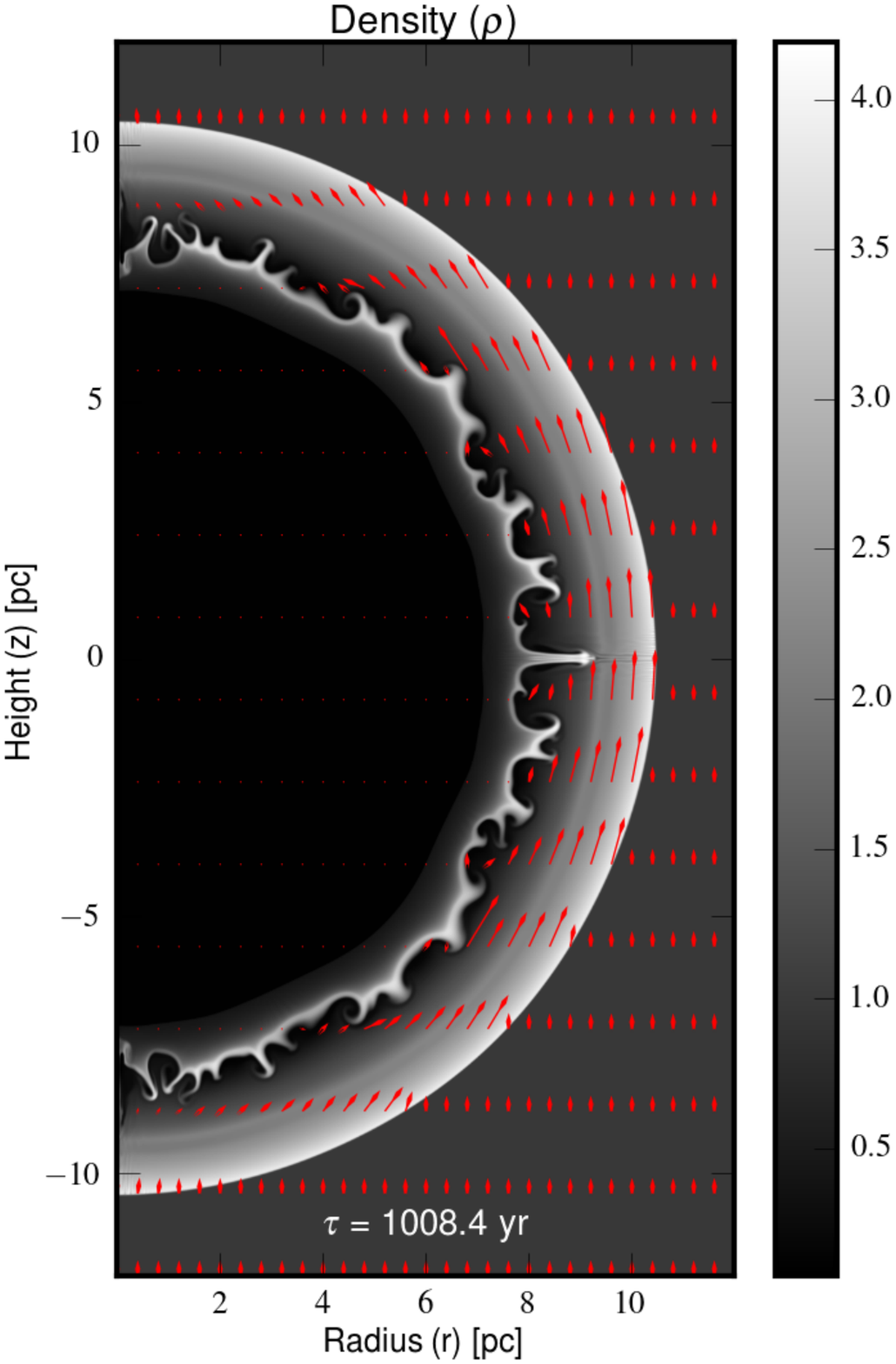}
  \figcaption{Evolution of fluid density at time $\tau \approx 1008$ yr  along with magnetic field vectors shown as \textit{red} arrows} 
\label{fig:sn1006_sim}
\end{figure}
 
Figure \ref{fig:sn1006_sim} shows the fluid density for the SNR at time $\tau = 1008$ yr. 
The magnetic field is represented by red arrows. 
We see the formation of Rayleigh-Taylor instabilities at the contact wave. 
The forward spherical shock traverses across the magnetic fields thereby modifying its vertical alignment. 
Due to compression from the shock, the magnetic flux just ahead of the shock is also enhanced and follows the curved shock as evident from the magnetic field vectors (shown in red).   

\begin{figure*}
\centering

\includegraphics[width=2\columnwidth]{\fpath/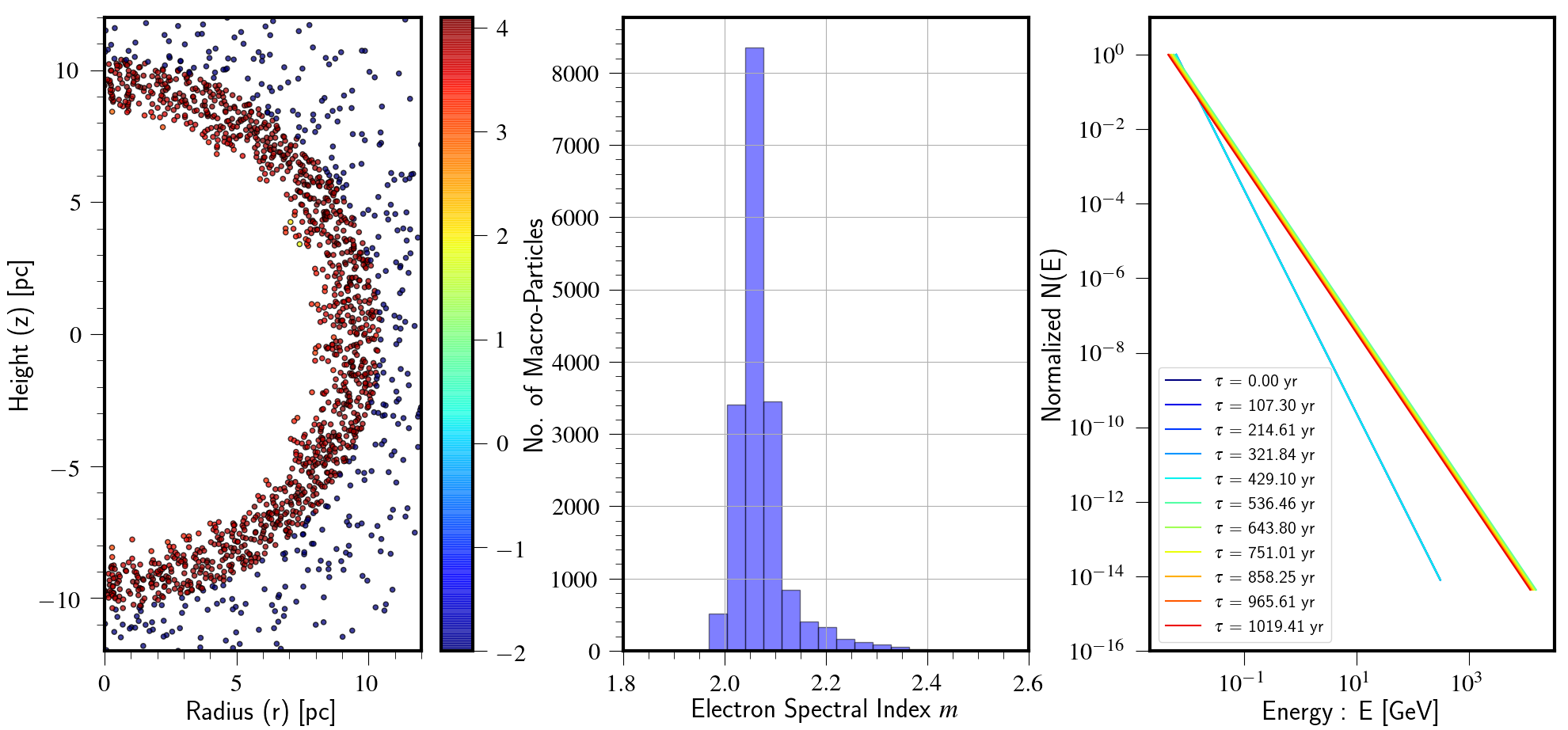}
\figcaption{(Left panel) Particle distribution at time $\tau = 1008$\,yr for the SN 1006 supernova remnant case. The colors represent the compression ratio due to shock while the negative values represent initial particles in the domain that have not interacted with shock. (Middle Panel) Histogram showing the electron spectral index $m$ for all the particles that have been shocked. (Right Panel) Evolution of spectral energy distribution for a single representative macro-particle.}
\label{fig:sn1006_pdistrib}
\end{figure*} 

A total of $2.5\times10^{4}$ macro-particles are randomly initialized in the ambient medium.  
To each of them we attach a scalar quantity "color" whose value is initially set to be -2  for all. 
However, as the simulation progresses in time, these macro-particles enter the shock and sample the compression ratio as described in Sec \ref{ssec:dsa}. 
The scalar "color" for each macro-particle is then replaced by the value of the compression ratio of the shock it experiences. This helps to separate the particle population for further diagnostics. 
The initial population of particles (for e.g. electrons) has a steep power-law spectral distribution with an index $m = 3$ covering a range  $E_{\rm min} \approx 0.63$ MeV to $E_{\rm max} \approx  0.31$ TeV.
The value inital $E_{\rm max}$ is set in accordance with the Larmor radius constraint ensuring that every single macro-particles consist of an initial energy distribution of micro-particles that are situated very nearby in physical space (within one grid cell).
This initial spectral distribution is evolved accounting for radiation losses due to adiabatic, synchrotron and IC effects. 

The macro-particle distribution (as scalar "color") at time $\tau = 1008$ yr  is shown in the left panel of Fig. \ref{fig:sn1006_pdistrib}. 
This distribution evidently shows that most  of the macro-particles have a compression ratio close to 4.0 indicating a strong adiabatic shock.  
For all the macro-particles that are shocked, we estimate the spectral energy index, $m$ using the shock compression ratio. 
We assume \textit{isotropic} injection whereby the spectral index depends solely on the compression ratio and is independent of the orientation of magnetic field with respect to the shock normal (see Eq. \ref{eq:qnonrel}). 
The histogram of the spectral energy index showing a distinct peak around $m \approx 2.05$  is shown in the middle panel of Fig. \ref{fig:sn1006_pdistrib}. 
Due to the skewness in the distribution, an arithmetic average of spectral energy index gives a value $\av{m} = 2.1$.  
This is equivalent to a spectral frequency index 0.55, a value that is slightly flatter as compared to the observed estimate of 0.6 at radio wavelengths. 
Note that the value of $m$ obtained here is immediately after the particle has traversed the shock. 
However, the subsequent evolution in a magnetized environment will result in radiative losses due to adiabatic expansion and synchrotron and IC losses which will effectively steepen the spectrum specially at very high energies. 
The spectral evolution is shown in the right panel of Fig. \ref{fig:sn1006_pdistrib} for a representative single macro-particle. 
This macro-particle experiences the shock around 540 years and its spectral energy distribution is flattened by the shock and also extended to higher energies.
{\blue The rapid spectral change is, once again, caused by our ``instantanous'' steady state approach to DSA.
The maximum energy obtained after the diffusive shock acceleration is $\approx 16$ TeV.
Such an estimate is a factor 2.5 times smaller than the upper limit obtained by fitting the electron spectra from young supernova remnant SN1006 \citep{Reynolds:1999} assuming a magnetic field of $10\;\mu$G.}
As the shock passes, the losses due to adiabatic expansion are evident from a uniform downward shift over time. 
{\blue Losses due to synchrotron cooling are insignificant due to cooling time being larger than evolution time for electrons with few TeV energies for the field strengths of the order of $10\;\mu$G in our simulations. }

\subsection{Shocks in Relativistic Slab Jets}
%
%
The second application studies the particle acceleration at shocks in a two-dimensional relativistic slab jet.

The initial condition consists of a cartesian domain having a spatial extends of (0, $D = 10\pi a$) and (-D/2, D/2) along the $x$ and $y$ plane respectively. 
The domain is discretized with $384^2$ grid cells. 
The slab jet is centered at y=0 and has a vertical extent of length $a = 200 pc$ on both sides of the central axis. 
The slab jet has a flow velocity given by a bulk Lorentz factor $\gamma = 5$ along the $x$ axis while the ambient medium is static.
In order to avoid excitation of random perturbation due to steep gradient in velocity at the interface we convolve the jet velocity with a smoothening function as described in \cite{Bodo:1995}. 
Additionally, a uniform magnetic field with a plasma $\beta = 10^3$ along the $x$ axis corresponding to a field strength of  $\approx 6 mG$ is introduced. 
As the main goal of this application is to model the  interaction of under-dense AGN jets with the ambient, we choose the jet with a density ratio of  $\eta = 10^{-2}$,
\begin{equation}\label{eq:slabjetrho}
  \frac{\rho(y, \eta)}{\rho_0}
   = \eta - (\eta - 1) {\rm sech}\left[\left(\frac{y}{a}\right)^{6}\right]
\end{equation}
where $\rho_0 = 10^{-4} \, \rm cm^{-3}$ is the density of jet on the central axis (i.e., $y = 0$). 
The jet is set to be in pressure equilibrium with the ambient i.e., $P_{\rm jet} = P_{\rm amb} = 1.5 \times 10^{-9} \, {\rm dyne \, cm}^{-2}$.
Periodic boundary conditions are imposed along the X axis and free boundary conditions are imposed at the top and bottom boundaries. 

This initial configuration at time $\tau = 0$ is perturbed with a functional form that can excite a wide range of modes. We perturb the $y$ component of the velocity using the anti-symmetric perturbation described by Eq. 2b of the \cite{Bodo:1995} paper. 
The amplitude of the perturbation is chosen to be 1\% of the initial bulk flow velocity. 
The wavelength of the fundamental mode is set equal  to the size of the computational domain along the $x$ direction, the corresponding wave number is $k_0 = 2\pi/D  = 0.2/a$. 
These perturbations grow with time as a consequence of the Kelvin Helmholtz instability, progressively steepen and develop into shocks. 
These oblique shocks are typically seen in AGN jets as the bulk jet flow interacts with surrounding ambient.

In order to study the effects of such shocks on the process of particle acceleration via DSA, we introduce 2 macro-particles per cell ($\sim 3 \times 10^{5}$ particles) at the initial time. 
 Macro-particles are initialized with a very steep initial power-law spectrum ($m = 15$ see Eq. \ref{eq:pdistini}) covering a wide spectral energy range of  8 orders of magnitude with $E_{\min} = 0.63$ keV to $E_{\max} = 63\,\rm TeV$ with 250 bins.
The initial number density of real particles is set to be ${\cal N}_0 = 10^{-4} \rho_0$.  
During the early stages of evolution when the shocks have yet to form, particles experience radiative losses due to synchrotron and IC processes. 
After the perturbations steepen to form shocks, particles are accelerated via DSA and their spectral distribution is modified as described in Sec. \ref{ssec:dsa}. 
The obliquity of magnetic field with respect to the shock normal is also accounted for in the estimate of  the post-shock electron spectral slope $q$ of the particle, by using Eqs. \ref{eq:qiso} and \ref{eq:qperp}.  
The free parameters used to determine the energy bounds of the shock modified spectral distribution are chosen as $\delta_n = 0.01$ and $\delta_e = 0.5$. 

\begin{figure}
\centering
\includegraphics[width=1\columnwidth]{\fpath/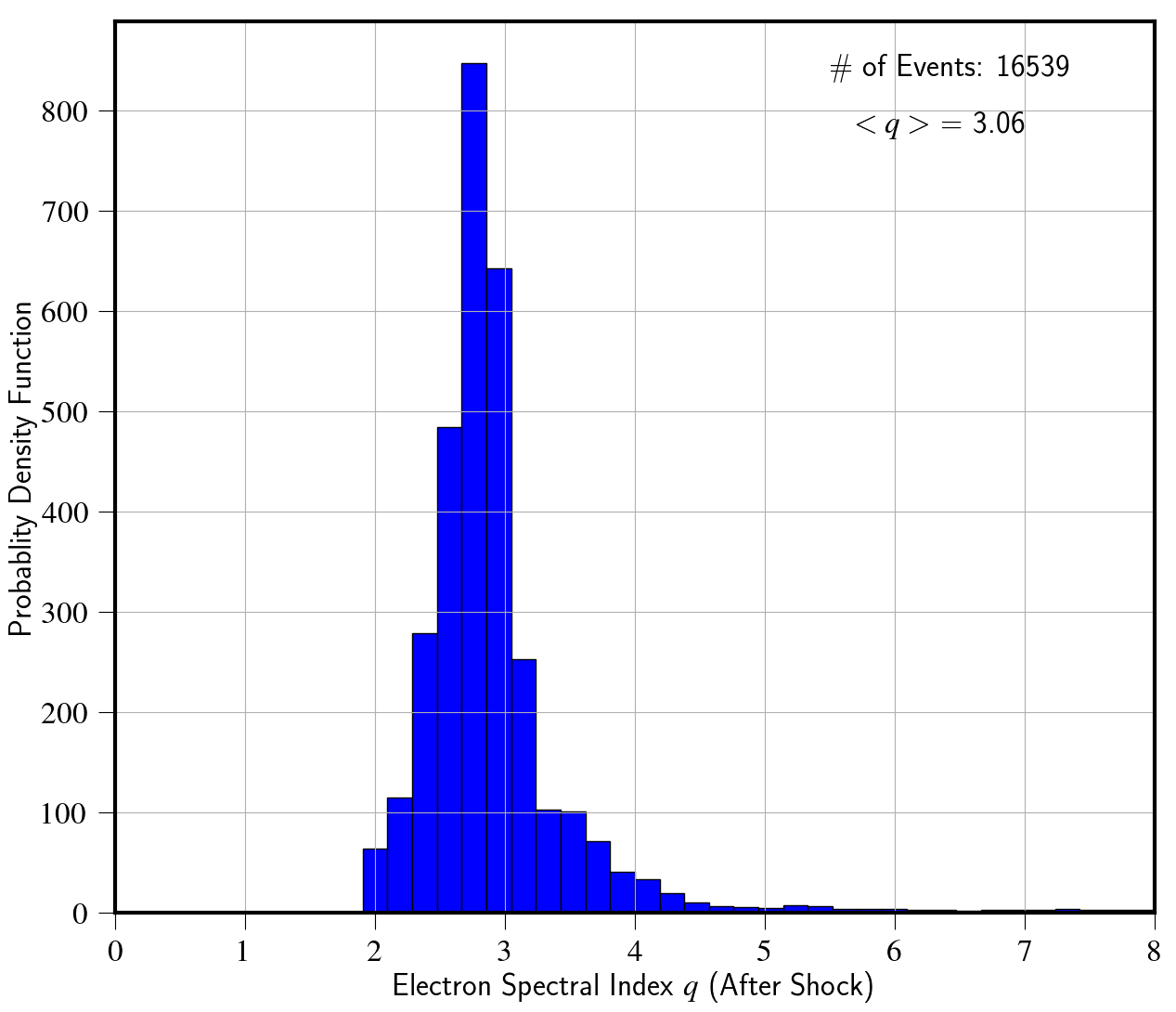}
\figcaption{Normalized PDF of the modified spectral slope $q$ as the particle crosses the shock during the evolution of slab jet until 0.3 Myr. }
\label{fig:slabjet_f1}
\end{figure}

During the simulation run of 0.3\,Myr, we record a total of 16539 events when the spectral distribution of the macro-particles is altered on passing through the shock.
The normalized probability distribution function (PDF) of the modified spectral slope $q$ is shown in Fig. \ref{fig:slabjet_f1}
The PDF shows a reasonable spread in the shock modified spectral slope $q$. We observe that the mean of the events of spectral modification results in a slope  $<q> \sim 3.1$
This spread arises due to our consistent approach of estimating the value of $q$ based on the compression ratio of the shock and the obliquity of the magnetic field with respect to shock normal.  
With our approach we relax the approximation of treating every shock  as a strong shock with a fixed spectral slope of $q = 2.23$ \citep{Mimica:2009,Fromm:2016} or $q = 2.0$ (\citealt{DeLaCita:2016}). 
The fixed choice of spectral index ($q \approx 2$) would result in an overestimate of the emissivity as the majority of shocks formed in our simulations have either lower strengths or are quasi-perpendicular resulting in a steeper spectral distribution.

\begin{figure*}
\centering
\includegraphics[width=1.95\columnwidth]{\fpath/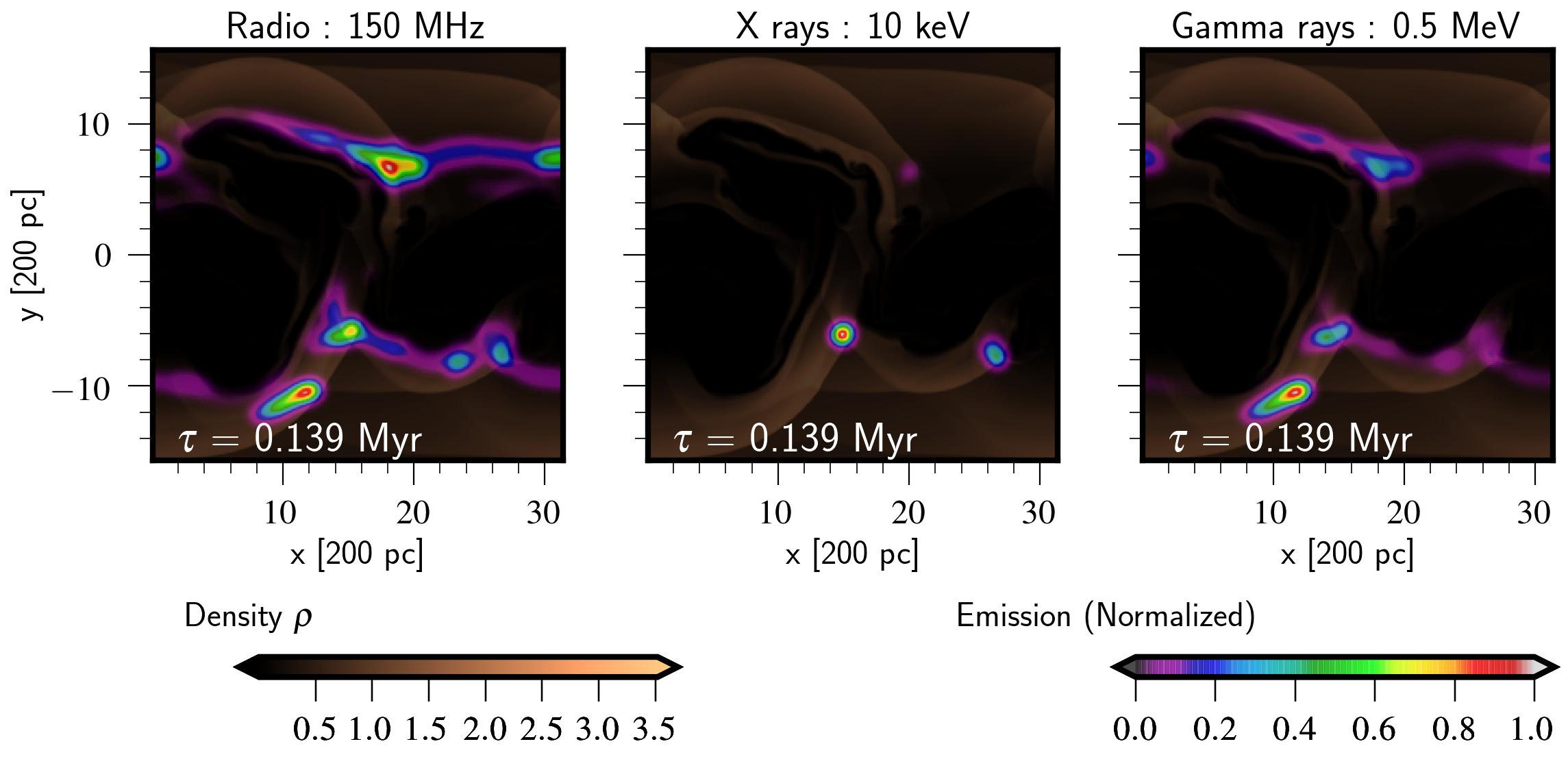}
\figcaption{Multi-wavelength emission signatures from slab jet simulation run at time $\tau = 0.137$\, Myr. Every panel shows the fluid density $\rho$ (as copper colors). 
The emissivities shown in each panel are obtained from instantaneous spectral distribution of particles and deposited on the grid. 
They are shown in spectral colors for three different observed frequencies viz., $\nu$ = 150 \, MHz (left), 10 \. keV (middle) due to synchrotron processes and 0.5 \, MeV (right) due to Inverse Compton. }
\label{fig:slabjet_f2}
\end{figure*}

We estimate the synchrotron emissivity $\mathcal{J}_{sy}(\nu, \hvec{n}_{los}, \vec{r})$, 
fractional polarization, $\Pi$ (see Eqs. \ref{eq:Jsy}, \ref{eq:Jpol}) and IC emissivity $\mathcal{J}_{IC}(\nu, \hvec{n}_{los}, \vec{r})$ (Eq. \ref{eq:JIC}) using the instantaneous spectral distribution for each macro-particle.
The above integral quantities for each macro-particle are then \textit{deposited} onto the fluid grid.  
The line of sight is chosen to be $\theta_{\rm obs} = 20^{\circ}$ with respect to the $z$-axis (pointing out of the plane).
The Gaussian convoluted normalized emissivity (with standard deviation $\sigma_{g}$ = 9)  is shown in the panels of Fig \ref{fig:slabjet_f2} for three different observed frequencies at time $\tau \sim 0.14$\,Myr. 
The left panel shows the emissivity at $\nu$ = 150 MHz in low frequency radio band using spectral colors. 
The emissivity for 10 keV X-ray energy is shown in the middle panel and the IC emissivity at an energy of 0.5\,MeV representing soft-gamma band is shown in the right panel. 
In each of these panels, we also show the fluid density $\rho$ in the background with \textit{copper} colors.
We observe a co-relation between high emissivity regions in the radio band with that of shocks formed as the jet interacts with the ambient medium. 
The X-ray emission at 10 keV is interesting and very distinct from the left and right panel. 
We observe X-ray emission as localized bright knots rather than a distributed emission in radio. The spots are associated with regions where there has been recent interactions of merging shocks as seen in the background fluid density.
The weak emission features observed in the right panel in the soft-gamma band are co-related with those seen in the left panel. No localized bright spots are observed for the emission at 0.5\,MeV.
This can be understood from the fact that, the same population of electrons responsible of the  production of the  low frequency radio emission also up-scatter CMB photons ($T_{CMB}(z)$ = 2.728\,K) to give rise to IC emission around the similar energy band.

\begin{figure*}
\centering
\includegraphics[width=2\columnwidth]{\fpath/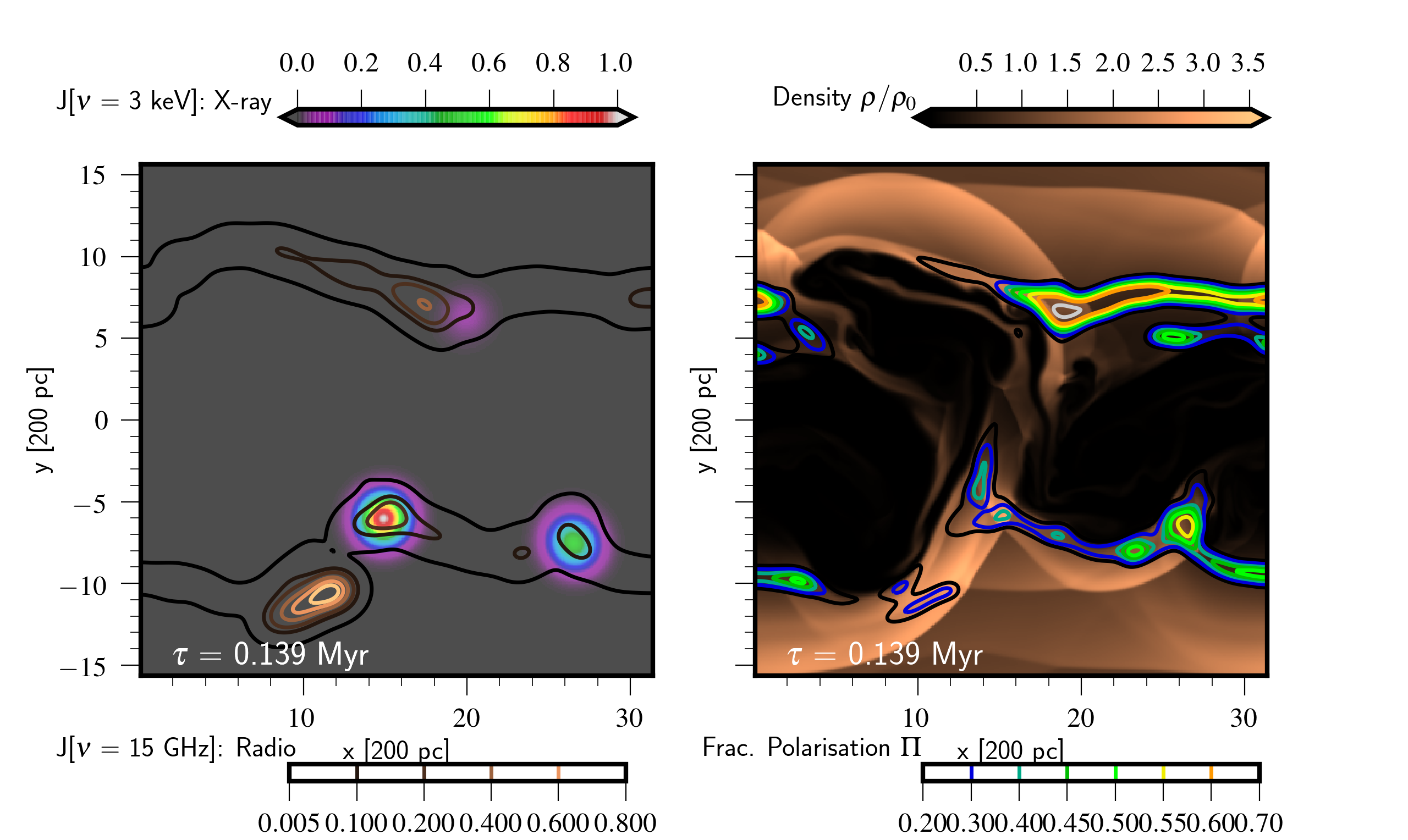}
\figcaption{Multi-wavelength emissivity map of the slab-jet at time $\tau = 0.137$\,Myr is shown in the \textit{left} panel. The colormap shows the Gaussian convoluted normalized X-ray (3 keV) emissivity and overlaid are the  contours for Gaussian convoluted normalized radio ($\nu = 15$\,GHz) emissivity. The \textit{right} panel shows the density of the flow in \textit{copper} color map at the same time and overlaid are the contours of the polarization degree whose value range from $\Pi = 20\%$ (black contours) to $\Pi = 70\%$ (gray contours). }
\label{fig:slabjet_f3}
\end{figure*}

To better compare the distinct nature of radio and X-ray synchrotron emission, we overlap the normalized X-ray emission corresponding to an energy of 3 keV with normalized radio ($\nu = 15$ GHz) contours in the left panel of Fig.\ref{fig:slabjet_f3}. 
The X-ray emission is convoluted with a beam that is 1.5 times broader than that used to obtain the radio contours.
Though our emissivity estimates from the slab jet are not integrated along the line of sight, we do see clear evidence of knotty emission in the X-ray bands that is offset from the radio peaks. 
The reason for this offset lies in the fact that they originate from different regions associated with the structure of oblique shocks. 
Radio emission is mainly forming due to large scale long lived shocks as the jet flow interacts with the ambient. 
Additionally, the radio electrons have a much longer synchrotron life time allowing them to produce bright emission in low frequencies. 
As the large scale forward moving shocks interact, they also result in the  formation of reverse shocks which eventually merge. 
Bright X-rays knots are produced  where such a recent merging of reverse shocks takes place and are short lived due to very short synchrotron cooling time of high energy electrons. 
Multi-wavelength observations of the kpc-scale jet in the powerful radio galaxy 3C 346 have shown signatures of an offset of about  0.8 kpc between the radio and X-ray emission  \citep{Worrall:2005, Dulwich:2009}. 
The synthetic emissivity map obtained from our simulations of oblique shocks is able to very well reproduce such offsets.

Additionally, the magnetic obliquity plays a crucial role in determining the spectral index and energy bounds of injected spectrum at shocks. 
The magnetic fields at oblique shocks typically become perpendicular to the jet flow therefore would result in steeper spectral slope. 
This can been understood from the distribution of fractional polarization shown in the right panel of Fig. \ref{fig:slabjet_f3}. 
We have overlaid contours (spectral colors) of $\Pi$ for radio band $\nu = 15 GHz$ on the \textit{copper} background of fluid density. 
The contour levels vary from 20\% (black) to 70\% (gray). Regions of high degree of polarization $> 50\%$  are seen at the merging large scale shocks indicating strong polarization of synchrotron emission at shocks.  
Multi-wavelength spectral studies of typical AGN jets like M87 and 3C 264 have shown evidences of X-ray synchrotron emission and harder spectral indices towards the edge of the jet \citep{Perlman:1999, Worrall:2005, Perlman:2010}.  
A consequence of this is presence of high degree of polarization at the edges of interface between the jet bulk flow and ambient medium. Optical and radio polarization studies in 3C 264 as well show a similar high degree $> 45\%$ close to edges \citep[e.g.,][]{Perlman:2006, Perlman:2010}.

Thus, our implementation of DSA at relativistic shocks for the case of slab jets shows similar qualitative features as observed for typical AGN jets. 
A one to one comparison with observed flux estimates will be taken up in subsequent paper using 3D RMHD jet simulations. 

\section{Discussion \& Conclusion}
\label{sec:summary}
%
%
%
We have presented a state-of-the-art hybrid framework for the PLUTO code that describes the spectral evolution of highly energetic particles by means of (mesh-less) Lagrangian macro-particles embedded in a classical or relativistic MHD fluid.
The main purpose of this work is that of including sub-grid micro-physical processes at macroscopic astrophysical scales where the fluid approximation is adequate.
While the MHD equations are integrated by means of standard Godunov-type finite-volume schemes already available with the code, macro-particles obey the relativistic cosmic-ray transport equation in the diffusion approximation.
Back reaction from particle to the fluid is not included and will be considered in forthcoming works.

The main features that characterize our hybrid framework are summarized below.
\begin{itemize}

\item
Lagrangian macro-particles follow fluid streamlines and embody a collection of actual physical particles (typically electrons) with a finite distribution in energy space.
For each macro-particle we solve, away from shocks, the cosmic ray transport equation in momentum (or energy space) to model radiation losses due to synchrotron, adiabatic expansion and inverse Compton effects based on local fluid conditions.
The transport equation is solved semi-analytically using the method of characteristics to update the energy coordinates in a Lagrangian discretization.

\item
In presence of magnetized shocks, we have described a novel technique to account for particle energization due to diffusive shock acceleration processes.
This involves sampling the local fluid quantities (such as velocity, magnetic field and pressure) in the upstream and downstream states to estimate the shock velocity along with the shock normal.
These quantities are critical to perform the transformation to the normal incidence frame where the compression ratio can then be calculated.
We have verified the validity of our shock-detection scheme by comparing it against theoretical estimates from 2D planar shocks.
The technique works also for curved as well as oblique shocks with very good accuracy.

The knowledge of the shock normal and of the local magnetic field direction enables us to include obliquity dependence in the estimate of the post-shock power-law index of the particle energy distribution.
In such a way our model is able to distinguish between quasi-parallel (more efficient) and quasi-perpendicular shocks (less efficient), the latter resulting in a steeper spectrum and depending on the amount of parameterized (unresolved) turbulence.
In both cases, the high energy cut-off is estimated consistently from the acceleration time scale derived without assuming equipartition but, rather, by considering particle diffusion along and across the magnetic field lines.

\item
The spectral distribution from each macro-particle is then further employed to compute observable such as emissivity and the degree of polarization due to synchrotron processes.
Numerical benchmarks involving a relativistically expanding shell have been used to demonstrate the accuracy of our implementation against theoretical expectation.
We adopt appropriate relativistic kinematic effects to estimate the observed degree of polarization and study its dependence on the viewing angles, $\theta_{\rm obs}$.
We observe that the value of polarization degree saturates for larger viewing angles. For $\gamma$-ray energies, we obtain $\Pi \approx 56\%$ for a power-law distribution with $m = 3$, smaller than the theoretical upper limit of 75\%.
This effect of depolarisation is consistent with values estimated by \cite{Lyutikov:2003}.
\end{itemize}

We have further applied our new framework to problems of astrophysical relevance  involving either classical MHD or relativistic magnetized shocks. 
Two examples have been proposed.
\begin{itemize}

\item \textbf{SN 1006}:
In the first application, we have studied diffusive shock acceleration and non-thermal emission in the context of supernova remnants with particular attention to SN1006.
Our study of particle acceleration at classical MHD shocks using axisymmetric SNR simulations has shown that the average spectral index for particles is around $m = 2.1$ consistent with values obtained for strong shocks. 
The maximum spectral energy of 20 TeV obtained for the magnetic  field of $ \sim 8 \mu$\,G is about a factor two times less than the upper limits obtained from fitting of observed spectra from SN 1006.

\item \textbf{Slab Jet}:
In the second application, we have investigated particle acceleration at shocks in a two-dimensional relativistic slab jet.
Unlike previous authors who employed a constant value for the spectral index of shocked particles, our method self-consistently determines the shock compression ratio and distinguishes between quasi parallel or quasi-perpendicular shocks.
This has shown to produce a considerable spreading in the electron spectral index (see Fig.\ref{fig:slabjet_f1}).
Also, we observe knotty emission features for X-ray energies and mis-aligned emissivity features indicating the effects of oblique shocks.
The polarization degree is also found to be larger at the jet/ambient interface, in agreement with radio and optical polarisation signatures from 3C 264 \cite{Perlman:2010}.
\end{itemize} 

{\blue
Forthcoming extensions of this work will aim at relaxing some of the simplifying assumptions adopted here.
In particular, efforts will be taken to: i) incorporate energy dependence in the free parameter $\eta$ for quasi-perpendicular relativistic shocks along with magnetic field amplification through feedback, ii) include macro-particle backreaction on the underlying fluid which can also account for modifications in the shock structure \citep{Blasi:2002}, iii) extend our framework to also include spectral evolution of protons with an aim to compare leptonic and hadronic emission from jets. 
The ultimate goal of this framework would be to model multi-wavelength emission from AGN jets by using three dimensional simulations.}

\section*{Acknowledgement}
We would like to thank the referee for providing valuable comments that has helped to improve the manuscript significantly. 
BV would like to thank Petar Mimica, Salvatore Orlando and Dipanjan Mukherjee for useful discussions.This work was supported by the funds provided by University of Torino. BV would also like to thank travel support by IAU for attending symposium on \textit{Perseus in Sicily : from black hole to cluster outskirts} in Noto where part of this work was presented. 


\bibliographystyle{apj}
\bibliography{paper}

\begin{thebibliography}{}
\expandafter\ifx\csname natexlab\endcsname\relax\def\natexlab#1{#1}\fi
\providecommand{\url}[1]{\href{#1}{#1}}

\bibitem[{{Achterberg} {et~al.}(2001){Achterberg}, {Gallant}, {Kirk}, \&
  {Guthmann}}]{Achterberg:2001aa}
{Achterberg}, A., {Gallant}, Y.~A., {Kirk}, J.~G., \& {Guthmann}, A.~W. 2001,
  \mnras, 328, 393

\bibitem[{{Aloy} {et~al.}(2000){Aloy}, {G{\'o}mez}, {Ib{\'a}{\~n}ez},
  {Mart{\'{\i}}}, \& {M{\"u}ller}}]{Aloy:2000}
{Aloy}, M.-A., {G{\'o}mez}, J.-L., {Ib{\'a}{\~n}ez}, J.-M., {Mart{\'{\i}}},
  J.-M., \& {M{\"u}ller}, E. 2000, \apjl, 528, L85

\bibitem[{{Bai} {et~al.}(2015){Bai}, {Caprioli}, {Sironi}, \&
  {Spitkovsky}}]{Bai:2015}
{Bai}, X.-N., {Caprioli}, D., {Sironi}, L., \& {Spitkovsky}, A. 2015, \apj,
  809, 55

\bibitem[{{Ballard} \& {Heavens}(1991)}]{Ballard:1991}
{Ballard}, K.~R., \& {Heavens}, A.~F. 1991, \mnras, 251, 438

\bibitem[{Birdsall \& Langdon(2004)}]{Birsdall_and_Langdon.2004}
Birdsall, C., \& Langdon, A. 2004, Plasma Physics via Computer Simulation,
  Series in Plasma Physics and Fluid Dynamics (Taylor \& Francis).
\newblock \url{https://books.google.it/books?id=S2lqgDTm6a4C}

\bibitem[{{Blandford} \& {K{\"o}nigl}(1979)}]{Blandford:1979}
{Blandford}, R.~D., \& {K{\"o}nigl}, A. 1979, \apj, 232, 34

\bibitem[{{Blandford} \& {Ostriker}(1978)}]{Blandford:1978aa}
{Blandford}, R.~D., \& {Ostriker}, J.~P. 1978, \apjl, 221, L29

\bibitem[{{Blasi}(2002)}]{Blasi:2002}
{Blasi}, P. 2002, Astroparticle Physics, 16, 429

\bibitem[{{Bodo} {et~al.}(1995){Bodo}, {Massaglia}, {Rossi}, {Rosner},
  {Malagoli}, \& {Ferrari}}]{Bodo:1995}
{Bodo}, G., {Massaglia}, S., {Rossi}, P., {et~al.} 1995, \aap, 303, 281

\bibitem[{{B{\"o}ttcher} \& {Dermer}(2010)}]{Boettcher:2010}
{B{\"o}ttcher}, M., \& {Dermer}, C.~D. 2010, \apj, 711, 445

\bibitem[{{Daldorff} {et~al.}(2014){Daldorff}, {T{\'o}th}, {Gombosi},
  {Lapenta}, {Amaya}, {Markidis}, \& {Brackbill}}]{Daldorff:2014}
{Daldorff}, L.~K.~S., {T{\'o}th}, G., {Gombosi}, T.~I., {et~al.} 2014, Journal
  of Computational Physics, 268, 236

\bibitem[{{de la Cita} {et~al.}(2016){de la Cita}, {Bosch-Ramon},
  {Paredes-Fortuny}, {Khangulyan}, \& {Perucho}}]{DeLaCita:2016}
{de la Cita}, V.~M., {Bosch-Ramon}, V., {Paredes-Fortuny}, X., {Khangulyan},
  D., \& {Perucho}, M. 2016, \aap, 591, A15

\bibitem[{{Del Zanna} {et~al.}(2006){Del Zanna}, {Volpi}, {Amato}, \&
  {Bucciantini}}]{DelZanna:2006}
{Del Zanna}, L., {Volpi}, D., {Amato}, E., \& {Bucciantini}, N. 2006, \aap,
  453, 621

\bibitem[{{Drury}(1983)}]{Drury:1983aa}
{Drury}, L.~O. 1983, Reports on Progress in Physics, 46, 973

\bibitem[{{Dulwich} {et~al.}(2009){Dulwich}, {Worrall}, {Birkinshaw},
  {Padgett}, \& {Perlman}}]{Dulwich:2009}
{Dulwich}, F., {Worrall}, D.~M., {Birkinshaw}, M., {Padgett}, C.~A., \&
  {Perlman}, E.~S. 2009, \mnras, 398, 1207

\bibitem[{{English} {et~al.}(2016){English}, {Hardcastle}, \&
  {Krause}}]{English:2016}
{English}, W., {Hardcastle}, M.~J., \& {Krause}, M.~G.~H. 2016, \mnras, 461,
  2025

\bibitem[{{Fromm} {et~al.}(2016){Fromm}, {Perucho}, {Mimica}, \&
  {Ros}}]{Fromm:2016}
{Fromm}, C.~M., {Perucho}, M., {Mimica}, P., \& {Ros}, E. 2016, \aap, 588, A101

\bibitem[{{Giacomazzo} \& {Rezzolla}(2006)}]{Giacomazzo:2006}
{Giacomazzo}, B., \& {Rezzolla}, L. 2006, Journal of Fluid Mechanics, 562, 223

\bibitem[{{Ginzburg} \& {Syrovatskii}(1965)}]{Ginzburg:1965}
{Ginzburg}, V.~L., \& {Syrovatskii}, S.~I. 1965, \araa, 3, 297

\bibitem[{{G{\'o}mez} {et~al.}(1997){G{\'o}mez}, {Mart{\'{\i}}}, {Marscher},
  {Ib{\'a}{\~n}ez}, \& {Alberdi}}]{Gomez:1997}
{G{\'o}mez}, J.~L., {Mart{\'{\i}}}, J.~M., {Marscher}, A.~P., {Ib{\'a}{\~n}ez},
  J.~M., \& {Alberdi}, A. 1997, \apjl, 482, L33

\bibitem[{{Gomez} {et~al.}(1995){Gomez}, {Marti}, {Marscher}, {Ibanez}, \&
  {Marcaide}}]{Gomez:1995}
{Gomez}, J.~L., {Marti}, J.~M.~A., {Marscher}, A.~P., {Ibanez}, J.~M.~A., \&
  {Marcaide}, J.~M. 1995, \apjl, 449, L19

\bibitem[{{Hardcastle} \& {Krause}(2014)}]{Hardcastle:2014}
{Hardcastle}, M.~J., \& {Krause}, M.~G.~H. 2014, \mnras, 443, 1482

\bibitem[{{Jokipii}(1987)}]{Jokipii:1987}
{Jokipii}, J.~R. 1987, \apj, 313, 842

\bibitem[{{Jokipii} \& {Parker}(1970)}]{Jokipii:1970}
{Jokipii}, J.~R., \& {Parker}, E.~N. 1970, \apj, 160, 735

\bibitem[{{Jones} {et~al.}(1999){Jones}, {Ryu}, \& {Engel}}]{Jones:1999}
{Jones}, T.~W., {Ryu}, D., \& {Engel}, A. 1999, \apj, 512, 105

\bibitem[{{Kardashev}(1962)}]{Kardashev:1962}
{Kardashev}, N.~S. 1962, \sovast, 6, 317

\bibitem[{{Keshet} \& {Waxman}(2005)}]{Keshet:2005aa}
{Keshet}, U., \& {Waxman}, E. 2005, Physical Review Letters, 94, 111102

\bibitem[{{Kirk} {et~al.}(2000){Kirk}, {Guthmann}, {Gallant}, \&
  {Achterberg}}]{Kirk:2000aa}
{Kirk}, J.~G., {Guthmann}, A.~W., {Gallant}, Y.~A., \& {Achterberg}, A. 2000,
  \apj, 542, 235

\bibitem[{{Kirk} \& {Reville}(2010)}]{Kirk:2010}
{Kirk}, J.~G., \& {Reville}, B. 2010, \apjl, 710, L16

\bibitem[{{Konigl}(1981)}]{Konigl:1981}
{Konigl}, A. 1981, \apj, 243, 700

\bibitem[{{Lichnerowicz}(1976)}]{Lichnerowicz:1976}
{Lichnerowicz}, A. 1976, Journal of Mathematical Physics, 17, 2135

\bibitem[{{Longair}(1994)}]{Longair:1994}
{Longair}, M.~S. 1994, {High energy astrophysics. Volume 2. Stars, the Galaxy
  and the interstellar medium.}

\bibitem[{{Lyutikov} {et~al.}(2003){Lyutikov}, {Pariev}, \&
  {Blandford}}]{Lyutikov:2003}
{Lyutikov}, M., {Pariev}, V.~I., \& {Blandford}, R.~D. 2003, \apj, 597, 998

\bibitem[{{Marcowith} {et~al.}(2016){Marcowith}, {Bret}, {Bykov}, {Dieckman},
  {O'C Drury}, {Lemb{\`e}ge}, {Lemoine}, {Morlino}, {Murphy}, {Pelletier},
  {Plotnikov}, {Reville}, {Riquelme}, {Sironi}, \& {Stockem
  Novo}}]{Marcowidth:2016}
{Marcowith}, A., {Bret}, A., {Bykov}, A., {et~al.} 2016, Reports on Progress in
  Physics, 79, 046901

\bibitem[{{Marscher}(1980)}]{Marscher:1980}
{Marscher}, A.~P. 1980, \apj, 235, 386

\bibitem[{{Mathews}(1971)}]{Mathews:1971}
{Mathews}, W.~G. 1971, \apj, 165, 147

\bibitem[{{Micono} {et~al.}(1999){Micono}, {Zurlo}, {Massaglia}, {Ferrari}, \&
  {Melrose}}]{Micono:1999}
{Micono}, M., {Zurlo}, N., {Massaglia}, S., {Ferrari}, A., \& {Melrose}, D.~B.
  1999, \aap, 349, 323

\bibitem[{{Mignone} {et~al.}(2007){Mignone}, {Bodo}, {Massaglia}, {Matsakos},
  {Tesileanu}, {Zanni}, \& {Ferrari}}]{plutocode:2007}
{Mignone}, A., {Bodo}, G., {Massaglia}, S., {et~al.} 2007, \apjs, 170, 228

\bibitem[{{Mignone} {et~al.}(2018){Mignone}, {Bodo}, {Vaidya}, \&
  {Mattia}}]{Mignone:2018}
{Mignone}, A., {Bodo}, G., {Vaidya}, B., \& {Mattia}, G. 2018, ArXiv e-prints,
  arXiv:1804.01946

\bibitem[{{Mignone} \& {McKinney}(2007)}]{Mignone:2007}
{Mignone}, A., \& {McKinney}, J.~C. 2007, \mnras, 378, 1118

\bibitem[{{Mignone} {et~al.}(2012){Mignone}, {Zanni}, {Tzeferacos}, {van
  Straalen}, {Colella}, \& {Bodo}}]{Mignone:2012}
{Mignone}, A., {Zanni}, C., {Tzeferacos}, P., {et~al.} 2012, \apjs, 198, 7

\bibitem[{{Mimica} \& {Aloy}(2012)}]{Mimica:2012}
{Mimica}, P., \& {Aloy}, M.~A. 2012, \mnras, 421, 2635

\bibitem[{{Mimica} {et~al.}(2009){Mimica}, {Aloy}, {Agudo}, {Mart{\'{\i}}},
  {G{\'o}mez}, \& {Miralles}}]{Mimica:2009}
{Mimica}, P., {Aloy}, M.-A., {Agudo}, I., {et~al.} 2009, \apj, 696, 1142

\bibitem[{{Miniati}(2001)}]{Miniati:2001}
{Miniati}, F. 2001, Computer Physics Communications, 141, 17

\bibitem[{{Park} {et~al.}(2015){Park}, {Caprioli}, \& {Spitkovsky}}]{Park:2015}
{Park}, J., {Caprioli}, D., \& {Spitkovsky}, A. 2015, Physical Review Letters,
  114, 085003

\bibitem[{{Parker}(1965)}]{Parker:1965}
{Parker}, E.~N. 1965, \planss, 13, 9

\bibitem[{{Perlman} {et~al.}(1999){Perlman}, {Biretta}, {Zhou}, {Sparks}, \&
  {Macchetto}}]{Perlman:1999}
{Perlman}, E.~S., {Biretta}, J.~A., {Zhou}, F., {Sparks}, W.~B., \&
  {Macchetto}, F.~D. 1999, \aj, 117, 2185

\bibitem[{{Perlman} {et~al.}(2006){Perlman}, {Padgett}, {Georganopoulos},
  {Sparks}, {Biretta}, {O'Dea}, {Baum}, {Birkinshaw}, {Worrall}, {Dulwich},
  {Jester}, {Martel}, {Capetti}, \& {Leahy}}]{Perlman:2006}
{Perlman}, E.~S., {Padgett}, C.~A., {Georganopoulos}, M., {et~al.} 2006, \apj,
  651, 735

\bibitem[{{Perlman} {et~al.}(2010){Perlman}, {Padgett}, {Georganopoulos},
  {Worrall}, {Kastner}, {Franz}, {Birkinshaw}, {Dulwich}, {O'Dea}, {Baum},
  {Sparks}, {Biretta}, {Lara}, {Jester}, \& {Martel}}]{Perlman:2010}
---. 2010, \apj, 708, 171

\bibitem[{{Porth} {et~al.}(2011){Porth}, {Fendt}, {Meliani}, \&
  {Vaidya}}]{Porth:2011}
{Porth}, O., {Fendt}, C., {Meliani}, Z., \& {Vaidya}, B. 2011, \apj, 737, 42

\bibitem[{{Reynolds} \& {Keohane}(1999)}]{Reynolds:1999}
{Reynolds}, S.~P., \& {Keohane}, J.~W. 1999, \apj, 525, 368

\bibitem[{{Schneiter} {et~al.}(2010){Schneiter}, {Vel{\'a}zquez}, {Reynoso}, \&
  {de Colle}}]{Schneiter:2010}
{Schneiter}, E.~M., {Vel{\'a}zquez}, P.~F., {Reynoso}, E.~M., \& {de Colle}, F.
  2010, \mnras, 408, 430

\bibitem[{{Schwartz}(1998)}]{Schwartz:1998}
{Schwartz}, S.~J. 1998, ISSI Scientific Reports Series, 1, 249

\bibitem[{{Sironi} {et~al.}(2015){Sironi}, {Keshet}, \&
  {Lemoine}}]{Sironi:2015}
{Sironi}, L., {Keshet}, U., \& {Lemoine}, M. 2015, \ssr, 191, 519

\bibitem[{{Sironi} \& {Spitkovsky}(2009)}]{Sironi:2009}
{Sironi}, L., \& {Spitkovsky}, A. 2009, \apj, 698, 1523

\bibitem[{{Sironi} {et~al.}(2013){Sironi}, {Spitkovsky}, \&
  {Arons}}]{Sironi:2013aa}
{Sironi}, L., {Spitkovsky}, A., \& {Arons}, J. 2013, \apj, 771, 54

\bibitem[{{Skilling}(1975)}]{Skilling:1975}
{Skilling}, J. 1975, \mnras, 172, 557

\bibitem[{{Summerlin} \& {Baring}(2012)}]{Summerlin:2012}
{Summerlin}, E.~J., \& {Baring}, M.~G. 2012, \apj, 745, 63

\bibitem[{{Takamoto} \& {Kirk}(2015)}]{Takamoto:2015aa}
{Takamoto}, M., \& {Kirk}, J.~G. 2015, \apj, 809, 29

\bibitem[{{Taub}(1948)}]{Taub:1948}
{Taub}, A.~H. 1948, Physical Review, 74, 328

\bibitem[{{Tregillis} {et~al.}(2001){Tregillis}, {Jones}, \&
  {Ryu}}]{Tregillis:2001}
{Tregillis}, I.~L., {Jones}, T.~W., \& {Ryu}, D. 2001, \apj, 557, 475

\bibitem[{Vaidya {et~al.}(2016)Vaidya, Mignone, Bodo, \&
  Massaglia}]{Vaidya:2016}
Vaidya, B., Mignone, A., Bodo, G., \& Massaglia, S. 2016, Journal of Physics:
  Conference Series, 719, 012023.
\newblock \url{http://stacks.iop.org/1742-6596/719/i=1/a=012023}

\bibitem[{{van Marle} {et~al.}(2018){van Marle}, {Casse}, \&
  {Marcowith}}]{vanMarle:2017}
{van Marle}, A.~J., {Casse}, F., \& {Marcowith}, A. 2018, \mnras, 473, 3394

\bibitem[{{Webb}(1989)}]{Webb:1989}
{Webb}, G.~M. 1989, \apj, 340, 1112

\bibitem[{{Webb} \& {Gleeson}(1979)}]{Webb:1979}
{Webb}, G.~M., \& {Gleeson}, L.~J. 1979, \apss, 60, 335

\bibitem[{{Worrall} \& {Birkinshaw}(2005)}]{Worrall:2005}
{Worrall}, D.~M., \& {Birkinshaw}, M. 2005, \mnras, 360, 926

\end{thebibliography}

\appendix

\section{Complete Analytic Solution for RMHD Shocks.}
\label{sec:AppA}
%
%
Here we describe the steps used to derive the analytic solution that completely describes 
the RMHD shock with arbitrary orientation of magnetic fields. 
For the tests of planar shocks described in this paper, the inputs are the pre-shock conditions (region where the particle is initialized) and the shock speed,  (treated as input parameter). Our aim is obtain the scalar and vector quantities in the post-shock region (where the particle moves on crossing the shock). Without the loss of generality we will assume here that the shock moves along the positive X axis. 

Let us denote input quantities, $\vec{U_{a}}$ in pre-shock region with \textit{sub-script a} and the unknown post-shock quantities, $\vec{U_{b}}$ with \textit{sub-script b}. In the lab frame, these quantities are related via the following jump condition across a fast magneto-sonic shock with speed $\vsh$,
\begin{equation}
\label{eq:jumpc}
\vsh \jump{\vec{U}} = \jump{\vec{F(U)}}.
\end{equation}
Here, $\jump{q} = q_{b} - q_{a}$ denotes the jump across the wave and $\vec{F(q)}$ is the flux for any quantity $q$.  The set of above jump conditions can be reduced to the following five positive-definite scalar invariant \citep{Lichnerowicz:1976, Mignone:2007} -
\begin{eqnarray}
\jump{J} &=& 0 \label{eq:jump1}\\
\jump{h\eta} &=& 0  \label{eq:jump2}\\	
\jump{\mathcal{H}} = \jump{\frac{\eta^{2}}{J^{2}} - \frac{b^{2}}{\rho^{2}}} &=& 0  \label{eq:jump3}\\
J^{2} + \frac{\jump{p + b^{2}/2}}{\jump{h/\rho}} & = & 0  \label{eq:jump4}\\
\jump{h^{2}} + J^{2}\jump{\frac{h^{2}}{\rho^{2}}} + 2\mathcal{H}\jump{p} + 2 \jump{b^{2} \frac{h}{\rho}} &=& 0  \label{eq:jump5},
\label{eq:lichnerocond}
\end{eqnarray}
where, $J = \rho\gamma_{s}\gamma(\vsh - \beta^{x})$ is the mass flux, $\gamma_{s}$ being the Lorentz factor of the shock and 
\begin{equation}
\eta = -\frac{J}{\rho}(\vec{v}\cdot\vec{B}) + \frac{\gamma_{s}}{\gamma}B^{x}.
\end{equation} 
The specific gas enthalpy $h$ is related to the gas pressure $p$ and density $\rho$ via an equation of state. The magnetic energy density, $b^{2}$  is related to the magnetic field $\vec{B}$ in lab frame as, 
\begin{equation}
|\vec{b}|^{2} =  \frac{|\vec{B}|^{2}}{\gamma^{2}} + (\vec{v}\cdot\vec{B})^{2}
\end{equation}

Following \cite{Mignone:2007}, we numerically solve the set of $3 \times 3$ non-linear equations \ref{eq:jump3}, \ref{eq:jump4} and \ref{eq:jump5} using the expression for the post-shock $\eta_{b} = \eta_{a}h_{a}/h_{b} $ from equation \ref{eq:jump2}. The solution of this closed set of equations, gives us the three unknown scalars viz., the gas pressure $p_{b}$, density $\rho_b$ and magnetic energy density $b^{2}_b$ in the post-shock region.

The next step in describing the shock completely is to estimate the post-shock vector quantities, i.e., velocities $\vec{\beta_{\rm b}} $ and magnetic fields $\vec{B_{b}}$. To estimate them, we use the exact Riemann solution for full set of RMHD equations \cite{Giacomazzo:2006}. In particular, we obtain the tangential components of the velocity ($\beta^{y}_{b}, \beta^{z}_{b}$) in the post-shock region using the expressions presented in \textit{Appendix A} of their paper.  These expressions relate the tangential velocity components to the pre-shock quantities and only the post-shock pressure, $p_{b}$. Further, using the estimated tangential velocity components, we obtain the normal velocity $\beta^{x}_{b}$ in the post-shock region using Eq. 4.25 in \cite{Giacomazzo:2006}. With the knowledge of post-shock velocity field, the magnetic fields in the post-shock region can be easily derived from the following jump conditions \cite{Giacomazzo:2006}, 
\begin{eqnarray}
\frac{J}{\gamma_s}\jump{\frac{B^{y}}{D}} + B^{x} \jump{\beta^{y}} &=& 0 \label{eq:jump6} \\
\frac{J}{\gamma_s}\jump{\frac{B^{z}}{D}} + B^{x} \jump{\beta^{z}} &=& 0 \label{eq:jump7},
\end{eqnarray}
where, $D = \rho \gamma$ is the proper gas density. 
Note that the magnetic field component normal to the shock front does not jump across the shock, i.e, $B^{x}_{a} = B^{x}_b$. The \textsc{Python} code written to derive the analytic solutions for RMHD shock conditions will be made available upon request from the author. 

\section{Frame Transformation to Normal Incidence Frame (NIF)}
\label{sec:frametrans}
%
%
In order to compute the spectral index of particle energy distribution as it passes the shock, one has to estimate the compression ratio in the shock rest frame. 
The compression ratio, $r$, is defined as the ratio of upstream to downstream velocities normal to the shock,   and since the mass flux is conserved across the shock, it is also equivalent to ratio of as the ratio of densities across the shock for non-relativistic MHD. 
\begin{equation}
\label{eq:jumpmhd}
r = \frac{\rho_{2}}{\rho_{1}} = \frac{\vec{v_{1}}\cdot \hvec{n}_{s}}{\vec{v_{2}}\cdot \hvec{n_{s}}},
\end{equation}
where the velocities $\vec{v_{1,2}}$ are obtained in shock rest frame which is defined in a unique way for non-relativistic MHD case. 

However, while treating relativistic MHD shocks, one can have multiple shock rest frames \citep{Ballard:1991, Summerlin:2012}. The \textit{Normal Incidence Frame} or NIF is the shock rest frame where the upstream velocity is normal to the shock front. The other often used shock rest frame in case of RMHD flows is the {\blue\textit{de Hoffmann-Teller}} Frame (HTF) wherein the upstream velocity and magnetic fields are aligned with the shock at rest. Since the HTF is usually defined for sub-luminal shocks and does not exist for super-luminal shocks, we choose to work with the NIF as our preferred shock rest frame. 

Given the shock speed, $\vsh$, normal to the shock direction, $\hvec{n}_{s}$ and both upstream and downstream states across the shock in the lab frame, we can transform to NIF in a two step process. The first step involves a Lorentz boost equal to shock velocity and along the direction of shock. Mathematically, any general four vector, $\vec{u}$ in lab frame is related to $\vec{u}^\prime$ in Lorentz boosted frame as follows,
\begin{equation}
\vec{u}^\prime = \mathcal{L}(\beta_{\rm bst}, \hvec{n}_{\rm bst})  \vec{u},
\end{equation}
where, the $\mathcal{L}$ is symmetric Lorentz boost operator.

For the first step, $\beta_{\rm bst} = \vsh$ and $\hat{n}_{\rm bst} = \hat{n}_{s}$. 
The second transformation requires another Lorentz boost to transform the intermediate primed frame of reference to obtain the NIF. 
In this case, the boost has to be in the transverse direction to the shock and with a boost velocity $\beta_{\rm bst} = v^{\prime}_{t}$, where, $v^{\prime}_{t}$ is the tangential velocity in the primed frame of reference. For two-dimensional tests with planar shocks propagating along the X axis, the tangential velocity is the velocity along Y axis obtained in the intermediate prime frame. 

With these two Lorentz boost, we obtain the quantities across the shock in NIF and then we can estimate the compression ratio as, 
\begin{eqnarray}
\label{eq:jumprmhd}
r & = & \frac{\vec{\beta}^{NIF}_{1} \cdot \hvec{n}^{NIF}}{\vec{\beta}^{NIF}_{2} \cdot \hvec{n}^{NIF}}\\
  & = & \frac{\rho_{2} \gamma^{NIF}_{2}}{\rho_{1} \gamma^{NIF}_{1}}
\end{eqnarray}

\end{document}